\documentclass[a4paper,11pt]{article}
\pdfoutput=1
\usepackage{jheppub}
\usepackage[T1]{fontenc} 
\usepackage{slashed,epsfig,amssymb,bm,graphicx,amsmath,amsfonts}
\usepackage{dcolumn}
\usepackage{mathrsfs}
\usepackage{color}
\newcommand{\be}{\begin{equation}}
\newcommand{\ee}{\end{equation}}
\newcommand{\ba}{\begin{eqnarray}}
\newcommand{\ea}{\end{eqnarray}}
\newcommand{\nn}{\nonumber}

\numberwithin{equation}{section}

\def\d{\delta}
\def\del{\partial}
\def\x{{\bf x}}
\def\y{{\bf y}}
\def\k{{\bf k}}
\def\q{{\bf q}}
\def\[{\left[}
\def\]{\right]}
\def\({\left(}
\def\){\right)}
\def\<{\langle}
\def\>{\rangle}
\def\O{\mathcal{O}}
\def\L{\mathcal{L}}
\def\H{\mathcal{H}}
\def\D{\mathcal{D}}
\def\F{\mathcal{F}}
\def\E{\mathcal{E}}
\def\eps{\epsilon}
\def\Mp{M_{\rm p}}
\def\kmin{k_{\rm min}}
\def\kmax{k_{\rm max}}
\def\rarr{\rightarrow}

\def\tzeta{\tilde{\zeta}}
\def\tHint{\tilde{\mathcal{H}}^{(\rm int)}}

\begin{document}

\title{Quantum Decoherence During Inflation from Gravitational Nonlinearities} 
\author{Elliot Nelson}
\affiliation{Perimeter Institute for Theoretical Physics, 31 Caroline St. N., Waterloo, ON, N2L 2Y5, Canada}

\emailAdd{enelson@perimeterinstitute.ca}

\date{\today}

\abstract{
We study the inflationary quantum-to-classical transition for the adiabatic curvature perturbation $\zeta$ due to quantum decoherence, focusing on the role played by squeezed-limit mode couplings. 
We evolve the quantum state $\Psi$ in the Schr\"odinger picture, 
for a generic cubic coupling to additional environment degrees of freedom.
Focusing on the case of minimal gravitational interactions, we find the evolution of the reduced density matrix for a given long-wavelength fluctuation by tracing out the other (mostly shorter-wavelength) modes of $\zeta$ as an environment.
We show that inflation produces phase oscillations in the wave functional $\Psi[\zeta(\x)]$, which suppress off-diagonal components of the reduced density matrix, leaving a diagonal mixture of different classical configurations. 
Gravitational nonlinearities thus provide a minimal mechanism for generating classical stochastic perturbations from inflation. 
We identify the time when decoherence occurs, which is delayed after horizon crossing due to the weak coupling, and find that Hubble-scale modes act as the decohering environment.
We also comment on the observational relevance of decoherence and its relation to the squeezing of the quantum state.
}

\maketitle

\tableofcontents

\section{Introduction}

A period of inflation in the early universe \cite{Guth:1980zm,Linde:1981mu,Albrecht:1982wi,Starobinsky:1982ee} provides a mechanism for stretching vacuum modes to super-Hubble scales
and generating large-scale, classical, stochastic perturbations \cite{Hawking:1982cz,Starobinsky:1982ee,Guth:1982ec}, drawn from a probability distribution $\rho$ over configurations.
These primordial perturbations have been measured precisely in the cosmic microwave background (CMB), most recently by the \textit{Planck} satellite \cite{Adam:2015rua,Planck:2015xua}.
The parameters of the distribution $\rho$ have been tightly constrained, indicating a distribution that is approximately Gaussian, scale-invariant, and adiabatic.

In the context of inflation this probability distribution is obtained from the quantum state of the inflaton and metric fluctuations: $\rho=|\Psi|^2$. There is thus an implicit assumption that a quantum measurement has been made, drawing a single realization for perturbations from this distribution, so that inflationary fluctuations may be treated as classical random variables.

Existing work on the quantum-to-classical transition \cite{Kiefer:2008ku} has focused largely on the squeezing of the quantum state for each mode during inflation \cite{Polarski:1995jg,Albrecht:1992kf,Grishchuk:1989ss,Grishchuk:1990bj,Guth:1985ya,Kiefer:1998qe}. 
As a result of this squeezing, access to non-commuting observables at the end of inflation is lost, and only the mod squared $|\Psi|^2$ -- which cannot be distinguished from a classical PDF -- is accessible to observation. Closely related is the fact that if we consider the quantum state for given mode $\zeta_\q$ of the scalar curvature perturbation, the spread in the conjugate variable $\dot{\zeta}_\q$ goes to zero after horizon crossing, $\sqrt{\<|\dot{\zeta}_\q|^2\>}\sim(k/aH)^2$, leaving the conjugate momentum inaccessible to observation. (In \S \ref{sec:zzz} we comment on the origin of this squeezing in the inflationary growth of the action, $\L\propto a(t)$, and resulting phase oscillations in the quantum state.)

However, this classicality does not address the question of how a measurement is made, so that the homogeneous quantum state - a coherent superposition of all field configurations - ``collapses'' to a particular stochastic realization of classical inhomogeneities. As is the case for any quantum measurement, this relies on a mechanism of quantum decoherence, and necessitates the presence of additional environment degrees of freedom that couple to long-wavelength perturbations as a measuring device. This process is controlled by the dynamics of the wave functional phase ${\rm arg}[\Psi](t)$, while $|\Psi|^2$ determines correlation functions.

A variety of different possible system-environment couplings have been considered for decoherence during inflation, including inflaton self-interactions \cite{Lombardo:2005iz,Martineau:2006ki}, coupling to short-wavelength inflaton fluctuations \cite{Lombardo:2005iz}, gravitational waves \cite{Calzetta:1995ys}, isocurvature or to additional fields \cite{Sakagami:1987mp,Brandenberger:1990bx,Prokopec:2006fc}, or entanglement between spatially separated Hubble volumes \cite{Sharman:2007gi}. We comment in more detail on how our approach compares to these works in \S \ref{sec:lit}.
Recently, it was shown \cite{Burgess:2014eoa} (see also \cite{Burgess:2006jn}) that long-wavelength fluctuations decohere rapidly provided they couple to an environmental sector with an interaction $H_{\rm int}=\int d^3\x a^3\zeta(\x)\mathcal{B}(\x)$, where $\zeta$ describes inflaton or curvature fluctuations, and $\mathcal{B}$ is a functional of high frequency modes satisfying certain conditions. Here we point out that such a coupling arises from general relativity (GR) and will thus be present in all inflation models.
As discussed further in \S \ref{sec:lit}, the present work builds on these previous works by focusing on the \textit{dynamics} of the decoherence process during inflation -- that is, in understanding at what point during inflation a given mode may be treated as a classical stochastic variable rather than a quantum oscillator.
Due to the approximate time translation and scale invariance of inflation, the dynamics of decoherence is the same for all scales; each mode becomes classical when it is stretched to a given physical scale, which happens later for shorter modes.

We will study quantum-to-classical behavior of the adiabatic curvature perturbation $\zeta$ -- the scalar mode active during inflation, which is directly related to the Mukhanov variable $u$ -- due to \textit{interactions that arise solely from the nonlinearity of GR}.
(A similar computation was carried out in \cite{Franco:2011fg} to study decoherence from the gravitational coupling of scalar and tensor modes using the decoherence functional framework.)
These interactions were originally obtained by expanding the Einstein-Hilbert action along with the action for a scalar field inflaton with a slow-roll potential \cite{Maldacena:2002vr}. However, the scalar mode $\zeta$ and its gravitational couplings arise in all inflation models; $\zeta$ is the Goldstone boson that arises from the breaking of time-translation invariance in the quasi- de Sitter background \cite{Cheung:2007st,Cheung:2007sv}. While other interactions may be stronger sources for decoherence, only gravitational interactions are guaranteed to be present, and thus, we will see, provide a minimal mechanism for decoherence. This validates the use of the quantum state as a classical probability distribution,
\be
\rho[\zeta(\x)] = |\Psi[\zeta(\x)]|^2,
\ee
which is used to calculate post-inflationary correlation functions.

We will first discuss a generic interaction which nonlinearly couples curvature fluctuations to an environment, and will focus on the specific case of the minimal gravitational self-interaction, Eq. \eqref{coupling}.
Working in the Schr\"odinger picture, for which the time dependence is captured in the state $|\Psi(t)\>$, we will see that (as sketched in \S \ref{sec:infl_meas_modes}) decoherence arises from rapid oscillations of the phase of the wave functional $\Psi[\zeta(\x)]$, due to the growth of the interacting part of the action, $\L_{\rm int}\propto a(t)$. As a result of this WKB classicality, the initially pure quantum state for each mode evolves into a classical mixture. \textit{We will also compute the time $t_{\rm deco}(k)$ at which decoherence occurs for a given mode, and identify the scale $k_{\rm env}$ of the environment modes which are the leading source of decoherence.}

The plan of the paper is as follows.
In the remainder of the introduction, we review quantum decoherence and describe the specific case of nonlinearities (mode coupling) during inflation, and give an overview of our results.
We then begin in \S \ref{sec:setup} by setting up the problem in the context of the inflating background and presenting the framework for the nonlinear coupling of $\zeta$ to an environment, denoted as $\E$.
In \S \ref{sec:WFev} we compute the evolution of the wave functional for a generic cubic coupling.
In \S \ref{sec:dec_factor} we trace out the environment modes and compute the evolution of off-diagonal components of the reduced density matrix for $\zeta$, in terms of the evolving wave functional $\Psi[\zeta,\E]$.
Then in \S \ref{sec:zzz} we focus on the specific case where the environment is (predominantly short-wavelength) modes of $\zeta$ itself, coupled via gravitational self-interactions (which are present in all inflationary models); we discuss the relevance of extreme IR modes in \S \ref{sec:IR}, and of squeezed-limit consistency relations in \S \ref{sec:cons_rel}.
We end with a discussion in \S \ref{sec:disc}, commenting on previous studies of the quantum-to-classical transition, decoherence of gravitational waves, and the observational relevance of decoherence during inflation.

For notation, we will use the Fourier decomposition convention $X(\x)\equiv\int\frac{d^3\k}{(2\pi)^3}X_\k e^{i\k\cdot\x}$ for any quantity $X=\zeta,\E$, etc.

\subsection{Quantum Decoherence, Inflation, and Mode Coupling} 
\label{sec:infl_meas_modes}

Let us first consider the simple case of a single degree of freedom (such as a single quantum oscillator, or the spin or position of a particle) coupled to various other degrees of freedom which play the role of an environment.
We suppose that there exists a set of system basis states $\{|S_i\>\}$ for which an initial environment state $|\E(t_0)\>$ will respond through the influence of an interaction Hamiltonian as
\be\label{entangleSE}
|\E(t_0)\>|S_i(t_0)\>\rarr|\E_i(t)\>|S_i(t)\>.
\ee
That is, the system will remain in a basis state $|S_i(t)\>$ while the environment degrees of freedom respond to the state of the system by evolving into a conditional state $|\E_i(t)\>$. If the environment acts as an ideal measuring device, then
\be\label{Estates_ortho}
\lim_{t\rarr\infty}\<\E_i|\E_j\> = \d_{ij},
\ee
so the $|\E_i\>$ act as orthogonal ``pointer states'' which record the state of the system. An initially coherent superposition of system states will thus become entangled with the environment,
\be\label{entangleSE2}
|\E(t_0)\>\(\sum_i \alpha_i |S_i(t_0)\> \) \rarr \sum_i \alpha_i |\E_i(t)\>\otimes|S_i(t)\>.
\ee

The orthogonality of the conditional environment states is directly related to the off-diagonal components of the reduced density matrix for the system. In terms of the global density matrix or operator $\hat{\rho}=|\Psi\>\<\Psi|$ (here, $|\Psi\>$ is the state of the system$+$environment), the reduced density matrix for the system is
\be
\hat{\rho}_{\rm R} \equiv {\rm Tr}_\E \hat{\rho} \equiv \sum_I \<\E_I| \hat{\rho} |\E_I\>,
\ee
where the $|\E_I\>$ is some choice of environment basis states.
If we trace out the environment from a global state of the form Eq. \eqref{entangleSE2}, $|\Psi\>=\sum_i \alpha_i |\E_i\>\otimes|S_i\>$, we find that the matrix elements are
\be\label{rho_r}
\rho_{\rm R}(S_i,S_j) \equiv \<S_i|\hat{\rho}_{\rm R}|S_j\> = \alpha_i\alpha_j^* \sum_I\<\E_j|\E_I\>\<\E_I|\E_i\>
=\alpha_i\alpha_j^*\<\E_i|\E_j\>,
\ee
so if $\<\E_i|\E_j\> \approx \d_{ij}$, then the reduced density matrix is nearly diagonal, describing a mixed state:
\be\label{rho_diag}
\rho_{\rm R}(S_i,S_j) \approx |\alpha_i|^2 \d_{ij}.
\ee
The effectiveness of the environment as a measuring device, then, is captured in the suppression of off-diagonal elements of the reduced density matrix.
(In the classic example of the double slit experiment, the $S_i$ are paths of electrons or other particles through different slits, and the damping of the interference pattern between these paths results from coupling to an environment such as the electromagnetic field, which records path information.)
If the system starts in a coherent superposition, it will be \textit{decohered} through the entanglement with its environment as in Eq. \eqref{entangleSE2}; this decoherence is captured in the off-diagonal suppression.

In the context of inflation, we are interested in the scalar curvature perturbation $\zeta(\x)$, the scalar mode that is present (along with two graviton modes) during inflation due to breaking of the time-translation invariance of de Sitter space \cite{Cheung:2007st}.
Using the ADM formalism to decompose the metric into spatial and time components, we may define $\zeta$ as fluctuations in the induced spatial metric,
\be\label{g_ij}
g_{ij} = a^2(t)e^{2\zeta(\x,t)},
\ee
where we have omitted tensor modes. The curvature perturbation thus describes the amount of expansion at any point, so $\zeta\ll1$ describes the local amount of expansion in any region.

The basis states of the quantized field $\hat{\zeta}(\x)$ which we consider are the field eigenstates $|\zeta(\x)\>$, which are specified by a classical configuration $\zeta(\x)$ and satisfy $\hat{\zeta}(\x)|\zeta(\x)\>=\zeta(\x)|\zeta(\x)\>$ when acted on by the operator $\hat{\zeta}(\x)$ at any point in space. We will eventually assume that $\zeta$ starts in the Bunch-Davies vacuum, which is a Gaussian state and a coherent superposition of field eigenstates,
\be\label{zeta_coherent}
|\Psi_\zeta\> \sim \sum_{\zeta} \alpha_{\zeta(\x)}|\zeta(\x)\>,
\ee
where $\alpha_{\zeta(\x)}=\<\zeta(\x)|\Psi_\zeta\>$ is a Gaussian functional, which we parametrize for Gaussian states in Eq. \eqref{PsiG} below.

\begin{figure}
\begin{center}
\includegraphics[scale=0.5]{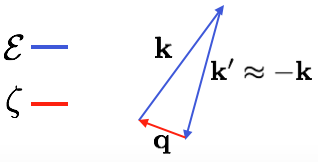}
\end{center}
\caption{A $\zeta\E^2$-type interaction couples modes of the curvature perturbation $\zeta$ to pairs of modes of an environment $\E$. Translation invariance requires that the momenta of three coupled modes satisfy $\k+\k'+\q=0$.}
\label{fig:triangle}
\end{figure}

We consider an additional field $\E(\x)$ as the environment. The dynamical degrees of freedom parametrized by $\E$ could come, for example, from isocurvature from an additional scalar field, or from tensor polarizations $h_+$ or $h_\times$. We will eventually restrict the system to a single Fourier mode of $\zeta$, and take $\E$ to be other modes of $\zeta$ itself. The leading nonlinearity is assumed to come from a cubic interaction $H_{\rm int}=\O(\zeta\E^2)$ which couples pairs of environment modes to modes of $\zeta$, as illustrated in Figure \ref{fig:triangle}.
This will result in a system-environment entanglement, described below in Eqs. \eqref{WF}-\eqref{PsiNG}.
Depending on the precise interaction, the environment field may respond to a classical configuration $\zeta(\x)$ by evolving into a conditional state $\left.\Psi_\E\right|_{\zeta(\x)}$, so that the entanglement of Eq. \eqref{entangleSE} takes the form
\be
|\Psi_\E\>|\zeta(\x)\> \rarr \(\left.|\Psi_\E\>\right|_{\zeta(\x)} \) |\zeta(\x)\>,
\ee
leading to the destruction of the coherent superposition of Eq. \eqref{zeta_coherent}.
The sum in Eq. \eqref{rho_r} takes the form of an integral over the environment field,\footnote{The $i$ and $j$ indices are replaced by configurations $\zeta$ and $\tzeta$.}
\be\label{env_int}
\rho_{\rm R}[\zeta(\x),\tzeta(\x)] =
\Psi_\zeta[\zeta(\x)]\Psi_\zeta^*[\tzeta(\x)]
\underbrace{\int\D\E\(\left.\Psi_\E[\E]\right|_{\zeta(\x)}\) \(\left.\Psi_\E[\E]\right|_{\tzeta(\x)}\)^*}_{\scalebox{1.0}{$=\Big(\<\Psi_\E|_{\zeta(\x)}\Big)\(|\Psi_\E\>_{\tzeta(\x)}\)$}}, 
\ee
which we compute below in Eq. \eqref{rho_env_int} and following.
The environment field makes records of the long-wavelength field $\zeta(\x)$ by carrying away phases,
\be\label{env_phase}
\left.\Psi_\E[\E(\x)]\right|_{\zeta(\x)}\propto e^{i\zeta\star\E\star\E\star({\rm Im}\F)},
\ee
where the $\star$'s denote a convolution as given in Eq. \eqref{PsiNG} below with a function $\F$ that we will compute. When the environment is sensitive as a measuring device, ${\rm Im}\F$ is large and the rapidly oscillating phase damps the integral in Eq. \eqref{env_int}, so that the off-diagonal ``interference pattern'' $\sim\Psi_\zeta[\zeta]\Psi_\zeta^*[\tzeta]$ is destroyed, and off-diagonal components vanish (see Figure \ref{fig:osc}).\footnote{We could not measure such an interference pattern anyways, since we can only make measurements in the field basis. In the double slit experiment, this would be analogous to only being able to measure which slit a particle went through, and not having access to measurements in the position basis on a distant screen.}

We will be especially interested in the \textit{squeezed limit} coupling between short scale environment modes and long wavelength $\zeta$ fluctuations shown in Figure \ref{fig:triangle}.


\subsection{Overview of Results}
\label{sec:results}

A quick sketch of the computation and results is as follows:

The cubic interaction\footnote{Here, $(\del\zeta)^2=\d^{ij}\del_i\zeta\del_j\zeta$ and $\del^2=\d^{ij}\del_i\del_j$. We are omitting other terms which are irrelevant for decoherence; see Eq. \eqref{zzzterms} and below, and footnote \ref{footnote_zzz_terms}. The choice of gauge here is one where matter or density fluctuations vanish on equal-time hypersurfaces, although the same interaction will be present in different gauges.} \cite{Maldacena:2002vr}
\be\label{coupling}
\L_{\rm int} = \eps(\eps+\eta) a(t)\zeta(\del_i\zeta)^2 + \dots
\ee
arises from the nonlinearity of GR (expanding the Einstein-Hilbert action to third order in the fluctuations), and is suppressed by the slow roll parameters $\eps\equiv-\dot{H}/H^2$ and $\eta\equiv\dot{\eps}/\eps H$
in comparison to the free Lagrangian for $\zeta$, Eq. \eqref{S2zeta} below.
Evolving the wave functional starting in the Bunch-Davies vacuum, we find that $\L_{\rm int}$ acts as a source for the non-Gaussian part of $\Psi$ (e.g. Eqs. \eqref{PsiNG}, \eqref{F_eom} below), and produces an exponentially growing phase in the superhorizon regime, 
\be\label{NGphase}
\text{phase of non-Gaussian part of  } \Psi[\zeta]\big|_t \ \propto \L_{\rm int}(t) \ \propto \ a(t). \hspace{1.5cm}
\ee

Treating short-wavelength modes $\zeta_S$ as an environment, schematically\footnote{Writing $\zeta=\zeta_L+\zeta_S$ as a sum of its long and short-wavelength parts, we will also have a $\zeta_S\del^i\zeta_S\del_i\zeta_L$ cross term, but since $\del_i\zeta_L\ll\del_i\zeta_S$, this term is subdominant.}
\be
\zeta(\del_i\zeta)^2 \rarr \zeta_L(\del_i\zeta_S)^2 \ \ \text{with} \ \ \zeta_S\equiv\E
\ee
and $\zeta_L$ a particular longer wavelength mode, and tracing them out as in Eq. \eqref{env_int} to find the (time-dependent) reduced density matrix for $\zeta_L$, we find that this phase is recorded by the short modes as described in \S \ref{sec:infl_meas_modes},
\be\label{psi_short_phase}
\Psi_{\zeta_S}(t)\big|_{\zeta_L}\propto \exp\[i\zeta_L(\del_i\zeta_S)^2\cdot(\text{small coupling})\cdot a(t) \]
\ee
as advertised in Eq. \eqref{env_phase}. The growing oscillations eventually suppress the integral over the environment and thus also off-diagonal elements, Eq. \eqref{env_int}, producing a mixed state. (See Eqs. \eqref{d_boxed_final}-\eqref{Gamma_zzz} and Figure \ref{fig:D(N)} for the main result, along with the definitions given in Eqs. \eqref{d_def}, \eqref{Gamma_def}.) So the interaction allows short modes to ``measure'' longer-wavelength modes, and thus decoheres or damps interferences of different long-wavelength configurations.

\begin{figure}
\begin{center}
\includegraphics[scale=0.55]{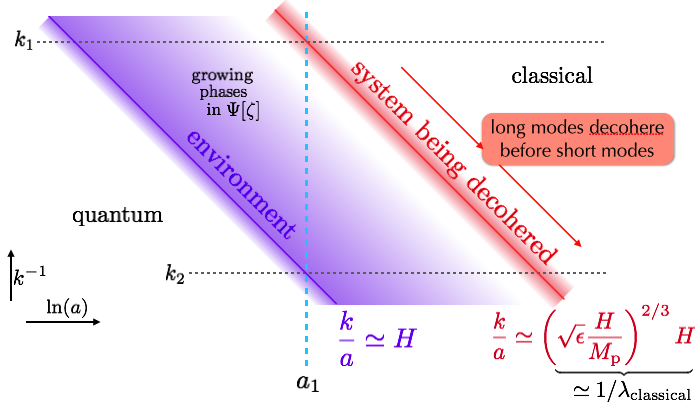}
\end{center}
\caption{\textbf{Decoherence from Mode Coupling during Inflation.} Here we show inverse wavenumber vs. $e$-foldings during inflation, $\ln a$. As modes redshift beyond the Hubble scale $k\simeq aH$, they become a sensitive as a decohering environment for longer modes. Once a super-Hubble mode reaches the physical wavelength $\lambda_{\rm classical}$, it is decohered, predominantly by shorter Hubble-scale modes. For example, $\zeta_{\k_1}$ becomes a classical stochastic fluctuation when $a(t)=a_1$ due to the decohering influence of Hubble-scale modes, $k_2\sim aH$. These modes are decohered in the same way several $e$-foldings later.}
\label{fig:QChorizon}
\end{figure}

The ``interaction strength'' of gravitational couplings -- the small coupling in Eq. \eqref{psi_short_phase} -- is proportional to the slow-roll parameters $(\eps+\eta)$ from Eq. \eqref{coupling}.
In the limit of $\eps,\eta\rarr0$ the interactions are negligible and each mode evolves independently, remaining in a pure state.
So when $(\eps+\eta)$ is nonzero but small, the weak coupling delays the pure-to-mixed transition described above:
While the non-Gaussian phase, Eq. \eqref{NGphase}, grows as $a(t)$ in the superhorizon regime, it only becomes $\O(1)$ in comparison to the Gaussian part and induces decoherence after an additional number of $e$-folds\footnote{There is also a dependence on $\eta$, which we gloss over here for simplicity, but include later.}
\be
N_{\rm deco}\approx\frac{1}{3}\ln\(\frac{1}{\eps^2}\frac{\eps \Mp^2}{H^2}\)
\ee
after the modes cross the horizon. Thus, fluctuations should not be viewed as classical stochastic variables upon horizon crossing, but only after inflating to a still larger scale. Once modes cross this ``classicality horizon'' and reach the scale
\be\label{class_horizon}
k_{\rm classical}\sim aH \( \frac{\eps H^2}{\Mp^2} \)^{1/3}, \ \ \ \text{or} \ \ \ \lambda_{\rm classical}\sim \frac{1}{H} \( \frac{\Mp^2}{\eps H^2} \)^{1/3},
\ee
as depicted in Figure \ref{fig:QChorizon}, they decohere rapidly. Specifically, the decoherence rate -- defined below in Eq. \eqref{Gamma_def} -- scales as the physical volume in Hubble units of the inflating region,
\be
\Gamma_{\rm decoherence}\propto (aH)^3,
\ee
and becomes $\O(1)$ when modes reach the scale given in Eq. \eqref{class_horizon}. At this point, off-diagonal components of the reduced density matrix are exponentially suppressed, with a decay time of order the Hubble time, that is within $\O(1)$ $e$-folds.

\begin{figure}
\begin{center}
\includegraphics[scale=0.35]{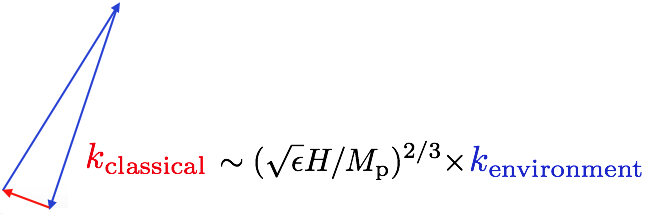}
\end{center}
\caption{Decoherence arises predominantly from ``squeezed limit'' configurations, for which the decohering mode couples to shorter environment modes, with $k_{\rm classical}/k_{\rm environment}\sim(\eps H^2/\Mp^2)^{1/3}$. ``Equilateral'' configurations with $k_{\rm environment}\lesssim k_{\rm classical}$ also contribute subdominantly.}
\label{fig:class_env_triangle}
\end{figure}

Shorter-wavelength modes become sensitive as an environment for superhorizon modes once they cross the Hubble scale, at which point they are the leading source of decoherence,
\be
k_{\rm environment}\sim aH.
\ee
This is a due to the fact that the interaction in Eq. \eqref{psi_short_phase} couples long-wavelength modes to the gradient of environment modes. The system-environment coupling that induces decoherence is dominated by the squeezed limit, as shown in Figure \ref{fig:class_env_triangle}. The long-wavelength background of $\zeta$ sets the local amount of expansion, which acts as a local time coordinate, and so short-scale modes act as a clock in the sense that they record or measure classical values for these background modes (see \S \ref{sec:cons_rel}).

The reader interested in a discussion of these results may skip to \S \ref{sec:disc}.

\section{Setup for the Problem}
\label{sec:setup}

\subsection{Review of Inflationary Perturbations}

We consider a quasi de Sitter phase of inflation, with a classical background metric given by
\be
ds^2 = -dt^2 + a^2(t) d\x^2,
\ee
where $a(t)\approx e^{Ht}$ is the scale factor. The slow-roll parameter $\eps\equiv-\dot{H}/H^2\ll1$ (overdots will denote $\frac{d}{dt}$) quantifies the deviation from exact de Sitter space, and is assumed to be constant up to higher order slow roll corrections.\footnote{We will work at leading order in $\eps$ throughout, treating the background as de Sitter, with $\eps$ only entering in the amplitude of fluctuations and their interactions.}
(We will not fix a specific matter source for inflation, but will assume that any matter present sources or is consistent with a quasi de Sitter background.)

The quadratic action for the adiabatic curvature perturbation $\zeta$ introduced in Eq \eqref{g_ij} is \cite{Brandenberger:1993zc,Maldacena:2002vr,Garriga:1999vw,Chen:2006nt}
\be\label{S2zeta}
S_{\rm free} = \frac{1}{2}\int d^4x \frac{2\eps \Mp^2}{c_s^2}a^3\(\dot{\zeta}^2 - c_s^2\frac{1}{a^2}(\del_i\zeta)^2\).
\ee
Here, $\Mp\equiv1/\sqrt{8\pi G}$ is the reduced Planck mass, and we have allowed a nontrivial speed of propogation, $0<c_s\leq1$.
We will treat $c_s$ as constant, and will focus on the case $c_s=1$, but see Appendix \ref{app:cs} for the more general case.

\subsection{Wave Functional for $\zeta$ and the Environment $\E$}

We will also include a collection of environmental degrees of freedom $\E$. In the example considered in \S \ref{sec:zzz}, $\E$ will be short-wavelength modes of $\zeta$ itself.
We will focus on the case where $\zeta$ is coupled to $\E$ through a cubic interaction that is linear in $\zeta$ and quadratic in $\E$. As show in Figure~\ref{fig:triangle}, an interaction of this type couples long-wavelength modes of $\zeta$ to a pair of shorter wavelength modes of $\E$, so that $\zeta$ may be sensitive to the high-energy bath of $\E$ modes.

Recall that in the Schr\"odinger picture, the time dependence of (equal-time) expectation values is captured in the wave functional evolution,
$\<\hat{\O}\>(t) = \<\Psi(t)|\hat{\O}|\Psi(t)\>,$ rather than in time-dependent operators, as in the Heisenberg picture. (In Appendix \ref{app:WF} we sketch the computation of the late-time three point function from the wave functional evolution.)
We also recall the field space or configuration space eigenstates $|\zeta,\E\>$ which satisfy
\be
\hat{\zeta}(\x)|\zeta(\x),\E(\x)\>=\zeta(\x)|\zeta(\x),\E(\x)\>, \ \ \ \hat{\E}(\x)|\zeta(\x),\E(\x)\>=\E(\x)|\zeta(\x),\E(\x)\>, \nn
\ee
and which are related to the wave functional as
\be
\Psi[\zeta,\E]\Big|_t = \<\zeta,\E|\Psi(t)\>. \nn
\ee

Assuming that the initial state $\Psi_0[\zeta,\E]\equiv\Psi[\zeta,\E]|_{t=t_0}$, evaluated at a time when all modes of interest are in the sub-Hubble regime ($k/a(t_0)H\gg1$),\footnote{In what follows we will send $t_0\rarr-\infty$ and choose the Bunch-Davies vacuum as the initial state at early times.} is Gaussian or close to Gaussian and that the couplings are weak, the evolved state will be nearly Gaussian throughout inflation, with a non-Gaussian part generated by the interactions. We parameterize the state as
\ba\label{WF}
\Psi[\zeta,\E] &=& \(\Psi^{(\zeta)}_G[\zeta]\) \times \(\Psi^{(\E)}_G[\E]\) \times \Big(\Psi_{NG}[\zeta,\E] \Big),
\ea
where the Gaussian pieces capture the free theory evolution for $\zeta$ and $\E$, and $\Psi_{NG}$ captures the non-Gaussian evolution from interactions. We assume that $\Psi$ is translationally and rotationally invariant, so that we can write the Gaussian parts as
\ba
\Psi^{(\zeta)}_G[\zeta](\tau) &=& N_\zeta(\tau)\exp\[-\int\frac{d^3\k}{(2\pi)^3}\zeta_\k\zeta_\k^*A_\zeta(k,\tau)\], \hspace{0.2cm} \nn \\
\Psi^{(\E)}_G[\E](\tau) &=& N_\E(\tau)\exp\[-\int\frac{d^3\k}{(2\pi)^3}\E_\k\E_\k^*A_\E(k,\tau)\].
\label{PsiG}
\ea
Here, $\tau\equiv-1/aH$ is conformal time, and $A_\zeta(k,\tau)$ is defined later in Eq. \eqref{AzetaBD}.
The real part of $A_{\zeta,\E}$ (required to be positive for a normalizable state) determines the equal-time two-point function; the imaginary part is related to the squeezing of the quantum state \cite{Polarski:1995jg} (see Eq. \eqref{AzetaBD} and following).
The non-Gaussian part, which is generated by the interaction, will be of the form
\ba\label{PsiNG}
\Psi_{NG}[\zeta,\E](\tau) = \exp\[\int_{\k,\k',\q}\E_\k\E_{\k'}\zeta_{\q}\F_{\k,\k',\q}(\tau)\],
\ea
where $\int_{\k,\k',\q}\equiv\int\frac{d^3\k}{(2\pi)^3}\frac{d^3\k'}{(2\pi)^3}\frac{d^3\q}{(2\pi)^3}(2\pi)^3\d^3(\k+\k'+\q)$ runs over momentum-conserving configurations\footnote{This is a consequence of assuming translation invariance of the initial state and action, so that $\Psi[\zeta(\x)]=\Psi[\zeta(\x+\y)]$ at all times.}. We take the complex kernel $\F$ to be symmetric in its first two arguments; the time dependence of its imaginary part will control the decoherence process.\footnote{We could also include a normalization factor in Eq. \eqref{PsiNG}, but the normalization of the state is only altered by the non-Gaussian part at $\O({\rm Re}\F)^2$; we will ignore effects from ${\rm Re}\F$ in \S \ref{sec:dec_factor}, as described there.}

We could also include a non-Gaussian piece coupling three environment modes, $\sim e^{\int\E\E\E}$. We ignore this part because if the environment is weakly self-interacting, it will not significantly affect the decoherence from system-environment interactions.

Note that the combination $\left.\Psi_\E\right|_{\zeta(\x)}\equiv\Psi^{(\E)}_G[\E(\x)] \Psi_{NG}[\zeta(\x),\E(\x)]$ may be viewed as the ``conditional'' wave function for the environment modulated by a fixed long-wavelength classical configuration $\zeta(\x)$. As seen in Eq. \eqref{env_int}, the degree of orthogonality between conditional wave functions for different configurations $\zeta$ (due to the influence of the environment) is equivalent to the suppression of off-diagonal components of the reduced density matrix for $\zeta$ (that is, of interferences between different configurations), which we compute in \S \ref{sec:dec_factor}.

The dynamics of $\Psi$ are captured in the complex functions $A_\zeta(k,\tau)$, $A_\E(k,\tau)$, and $\F_{\k,\k',\q}(\tau)$.  In \S \ref{sec:WFev} we will compute the evolution of $\F$ for an initial state specified by these functions at time $t_0$.

\section{Wave Functional Evolution for a Generic Cubic Interaction}
\label{sec:WFev}

In this section we evolve the wave functional $\Psi[\zeta,\E]$ during inflation for a generic cubic coupling between $\zeta$ and $\E$, in the Schr\"odinger picture. Then in \S \ref{sec:dec_factor} we will trace out the environment to obtain the dynamics of the reduced density matrix for $\zeta$.

The state evolves according to the Schr\"odinger equation
\be\label{Schro}
i\frac{d}{dt}\Psi[\zeta,\E] = H(t)\Psi[\zeta,\E],
\ee
We assume that the coupling can be treated perturbatively and that the initial state is at most perturbatively non-Gaussian, and thus break the Hamiltonian into the free (quadratic) Hamiltonians for $\zeta$ and $\E$, and interacting part,
\be
H[\zeta,\E] = H_\zeta[\zeta] + H_\E[\E] + H_{\rm int}[\zeta,\E],
\ee
but otherwise maintain generality. The free theory equations for the Gaussian parts of the state, Eq. \eqref{WF}, are
\ba
i\frac{d}{dt}\Psi^{(\zeta)}_G &=& H_\zeta(t) \Psi^{(\zeta)}_G, \label{Schr_free_zeta} \\
i\frac{d}{dt}\Psi^{(\E)}_G &=& H_\E(t) \Psi^{(\E)}_G, \label{Schr_free_E}
\ea
with solutions determined by $A_{\zeta,\E}(k,\tau(t))$ and $N_{\zeta,\E}(k,\tau(t))$ in Eq. \eqref{PsiG} \cite{Burgess:2014eoa,Polarski:1995jg}.\footnote{Here we assume an initial state that is at least nearly Gaussian, so that any non-Gaussian part may be treated perturbatively as part of $\Psi_{NG}$.} 
The free Hamiltonians are quadratic in $\zeta$ or $\E$ and their conjugate momenta, which act in the Schr\"odinger picture as
\be
\pi^{(\zeta)}_\k = -i\frac{\d}{\d\zeta_\k}, \hspace{0.7cm} \pi^{(\E)}_\k = -i\frac{\d}{\d\E_\k}.
\ee
We assume that the conjugate momenta appear in terms of the form
\be\label{Hkin}
H_\zeta^{(\rm kin)}=\frac{1}{2}\int\frac{d^3\k}{(2\pi)^3}f_\zeta(\tau)\pi^{(\zeta)}_\k\(\pi^{(\zeta)}_\k\)^*,
\hspace{0.5cm}
H_\E^{(\rm kin)}=\frac{1}{2}\int\frac{d^3\k}{(2\pi)^3}f_\E(\tau)\pi^{(\E)}_\k\(\pi^{(\E)}_\k\)^*.
\ee
Subtracting off Eqs. \eqref{Schr_free_zeta}-\eqref{Schr_free_E} from Eq. \eqref{Schro} and replacing $\frac{d}{dt}\rarr -H\tau\frac{d}{d\tau}$, we have
\ba
-H\tau\(\Psi^{(\zeta)}_G\Psi^{(\E)}_G\)i\frac{d}{d\tau}\Psi_{NG} &=& \Psi_G^{(\E)}\[H_\zeta^{(\rm kin)}\(\Psi_G^{(\zeta)}\Psi_{NG}\)\]_{\rm mixed} + \Psi_G^{(\zeta)}\[H_\E^{(\rm kin)}\(\Psi_G^{(\E)}\Psi_{NG}\) \]_{\rm mixed} \nn \\
&& + \Psi_{NG} H_{\rm int} \(\Psi^{(\zeta)}_G\Psi^{(\E)}_G\). \label{Schr_NG}
\ea
The notation is as follows:  In subtracting off Eqs. \eqref{Schr_free_zeta}-\eqref{Schr_free_E}, the only remaining terms are those for which the kinetic terms act partly on $\Psi_G^{(\zeta,\E)}$ and partly on $\Psi_{NG}$, indicated by ``mixed.''\footnote{We have also dropped terms in which factors of canonical momenta in $H_{\rm int}$ act partly on $\Psi_{NG}$. These terms will source quartic and higher order parts of the wave functional, which are more suppressed by the interaction strength and will not be captured correctly anyways when the action is truncated at third order in the fluctuations.}
Using Eqs. \eqref{PsiG}, \eqref{PsiNG}, and \eqref{Hkin}, we can write Eq. \eqref{Schr_NG} as
\be\label{F_eom_int}
0 = \int_{\k,\k',\q} \E_\k\E_{\k'}\zeta_{\q}\[i\F'_{\k,\k',\q}(\tau) + \alpha_{k,k',q}(\tau)\F_{\k,\k',\q}(\tau) + \frac{1}{H\tau} \tHint_{\k,\k',\q}(\tau) \].
\ee
Here we have denoted conformal time derivatives with a prime, and made two definitions. First,
\be\label{alpha_def}
\alpha_{k,k',q}(\tau) = \frac{1}{H\tau}\[f_\E(k,\tau)A_\E(k,\tau) + (k\rarr k')\] + \frac{1}{H\tau}f_\zeta(q,\tau)A_\zeta(q,\tau). \ee
Second, the interaction $H_{\rm int}$ generates non-Gaussianity in the state through the source\footnote{In general, the source $\tHint$ is nontrivially related to the Hamiltonian density $\H_{\rm int}$ because conjugate momenta introduce new $k$-dependence when acting on $\Psi_G$.} $\tHint$, obtained by acting $H_{\rm int}$ on the Gaussian wave functional $\Psi_G\equiv\Psi_G^{(\zeta)}\Psi_G^{(\E)}$,
\be\label{s_def}
(H_{\rm int}\Psi_G)\big|_\tau \equiv \[ \int_{\k,\k',\q}\E_\k\E_{\k'}\zeta_{\q} \(\tHint_{\k,\k',\q}(\tau)\) \] \Psi_G.
\ee
Eq. \eqref{F_eom_int} holds for all configurations $\{\zeta(\x),\E(\x)\}$, so we can set the integrand to zero,
\be\label{F_eom}
\F' - i\alpha(\tau)\F = i\tHint(\tau)/H\tau,
\ee
where we have suppressed the wavenumber labels on $\F$, $\alpha$, and $\tHint$.\footnote{Note that $\alpha$ and $\tHint$, like $\F$, can only be taken to be symmetric in the first two wavenumbers. However, in the case $\E=\zeta$ in \S \ref{sec:zzz} these quantities are completely symmetric.} We emphasize that these quantities are complex; our interest is primarily in the imaginary part of $\F$, which is relevant for decoherence. Eq. \eqref{F_eom} describes the non-Gaussianity in the state $\Psi$ generated by the nonlinear dynamics from the coupling $H_{\rm int}$ in Eq. \eqref{s_def}.
Assuming $\F(\tau_0)=0$,\footnote{If we allow for a nonzero correlation $\F_0$ in the initial state \cite{Agarwal:2012mq}, an additional term,
\be
\F(\tau_0)\exp\[i\int_{\tau_0}^\tau d\tau'\alpha(\tau')\], \nn
\ee
is present. In the examples we consider in \S \ref{sec:zzz} we will assume all modes start in the Bunch-Davies vacuum, and set $\F(\tau_0\rarr-\infty)$ to zero.}
the solution is
\be\label{F_solution}
\boxed{\F_{\k,\k',\q}(\tau) = i \int_{\tau_0}^\tau \frac{d\tau'}{H\tau'} \tHint_{\k,\k',\q}(\tau')\exp\[i\int_{\tau'}^\tau d\tau''\alpha_{k,k',q}(\tau'')\]. }
\ee
Here, the source $\tHint$ is determined by the interactions, and the exponential function is determined by the free theory evolution.

Let us consider the case where one or more terms in the interaction Hamiltonian cause the source $\tHint$ to grow at late times as a positive power of $a(t)$.
We are assuming that the system-environment coupling is weak enough that the probability $|\Psi|^2$ remains close to Gaussian, so ${\rm Re}\F$ should not grow at late times. This requires that the exponent -- and hence the imaginary part of the exponential -- becomes small as $\tau'\rarr\tau$. (See Eq. \eqref{alpha_zzz} and following for the case of gravitational self-interactions of $\zeta$.)
Hence, the non-Gaussian phase ${\rm Im}\F$ at late times is given by
\be\label{ImF_late}
\lim_{\tau\rarr0} {\rm Im}\F_{\k,\k',\q}(\tau) = i\int^\tau \frac{d\tau'}{H\tau'} \tHint_{\k,\k',\q}(\tau'). \hspace{1cm} (\tHint\propto a^p\propto \tau^{-p} \ \text{for } p>0)
\ee

In summary, Eq. \eqref{F_solution}, and in particular the case of the growing phase, Eq. \eqref{ImF_late}, describe the nonlinear evolution in the Schr\"{o}dinger picture, and can be used to compute the degree of decoherence from mode coupling. The nonlinear Schr\"{o}dinger picture evolution can also of course reproduce late-time correlation functions (see Appendix \ref{app:WF}).

\section{Integrating Out the Environment}
\label{sec:dec_factor}

In this section we will compute the reduced density matrix $\hat{\rho}_{\rm R}(\tau)$ for $\zeta$ -- here we will switch to conformal time -- obtained by tracing out the environment from the state $|\Psi(\tau)\>$, focusing on the late-time limit.
We will relate the dynamics of the non-linear part of the wave functional, Eq. \eqref{F_solution}, to the suppression of off-diagonal elements of $\hat{\rho}_{\rm R}$.
Then in \S \ref{sec:zzz} we will specify to the case of gravitational self-interactions of $\zeta$.

\begin{figure}
\begin{center}
\includegraphics[scale=0.5]{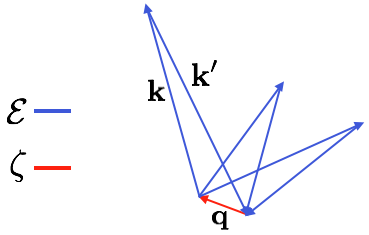}
\end{center}
\caption{Tracing out the environment $\E$ amounts to summing over pairs of Fourier modes $\E_\k$ that couple to a long wavelength curvature mode $\zeta_\q$. We will see in \S \ref{sec:zzz} that the dominant contribution to decoherence of $\zeta$ comes from Hubble-scale environment modes.}
\label{fig:sum}
\end{figure}

\begin{figure}
\begin{center}
\includegraphics[scale=0.43]{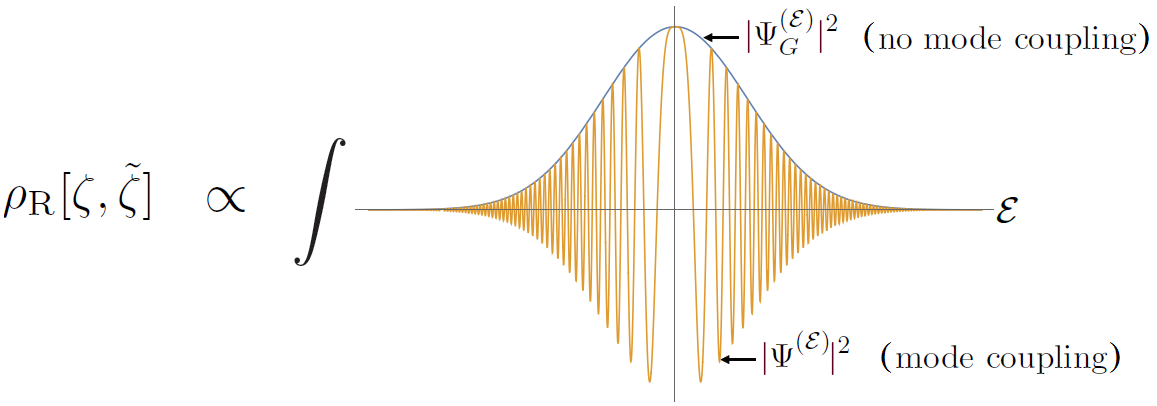}
\end{center}
\caption{An illustrative sketch of oscillations in the Gaussian integral over the environment in the presence of nonlinearities (mode couplings), due to a rapidly oscillating phase in Eq. \eqref{rho_env_int}.}
\label{fig:osc}
\end{figure}

The reduced density matrix for $\zeta$ is obtained from the global density matrix $\hat{\rho}\equiv|\Psi\>\<\Psi|$ by tracing or integrating over field configurations of the environment $\E$,
\be
\hat{\rho}_{\rm R}\equiv{\rm Tr}_\E\hat{\rho}\equiv\int\D\E\<\E|\hat{\rho}|\E\>.
\ee
Using Eq. \eqref{WF}, the reduced density matrix elements in the field basis, $\rho_{\rm R}[\zeta,\tzeta]\equiv\<\zeta|\hat{\rho}_{\rm R}|\tzeta\>$, are then given by 
\ba
\rho_{\rm R}[\zeta,\tzeta] &=& \int\D\E\Psi[\zeta,\E]\Psi^*[\tzeta,\E]  \label{rho_env_int} \\
&=& \Psi_G^{(\zeta)}[\zeta]\(\Psi_G^{(\zeta)}[\tzeta]\)^* \int\D\E|\Psi_G^{(\E)}|^2\exp\[\int_{\k,\k',\q}\E_\k\E_{\k'}\(\zeta_{\q}\F_{\k,\k',\q}+\tzeta_{\q}\F^*_{\k,\k',\q}\)\]. \nn
\ea
We will be interested in the amplitude of off-diagonal matrix elements as compared to the diagonal. We define the decoherence factor $D[\zeta,\tzeta]$, which quantifies the relative suppression from these oscillations, as
\ba\label{d_def}
D[\zeta,\tzeta] &\equiv& \frac{|\rho_{\rm R}[\zeta,\tzeta]|}{\sqrt{\rho_{\rm R}[\zeta,\zeta]\rho_{\rm R}[\tzeta,\tzeta]}} \\
&=& \int\D\E|\Psi_G^{(\E)}|^2\exp\[\int_{\k,\k',\q}\E_\k\E_{\k'}\(\zeta_{\q}\F_{\k,\k',\q}+\tzeta_{\q}\F^*_{\k,\k',\q}\)\].
\ea
In the second line we have ignored the real part of $\F$, which must be small if the interaction is small, so that the probability $|\Psi|^2$ will be close to Gaussian. The imaginary part ${\rm Im}\F$ may become very large, leading to a rapidly oscillating phase in the integrand (see Figure \ref{fig:osc}). We will focus on this case, for which ${\rm Im}\F$ is given by Eq. \eqref{ImF_late} at late times, and ignore the real part ${\rm Re}\F$, along with non-Gaussian contributions from environment self-interactions.

Note that we may write the decoherence factor as
\be\label{D_exp_value}
D[\zeta,\tzeta] =
\bigg\<\exp\[i \int_{\k,\k',\q}\E_\k\E_{\k'}\(\Delta\zeta_\q\){\rm Im}\F_{\k,\k',\q}\]\bigg\>_\E,
\ee
where $\Delta\zeta_\q\equiv\zeta_\q-\tzeta_\q$, and where $\<\O\>_\E\equiv\int\D\E|\Psi_G^{(\E)}|^2\O$ indicates the expectation value in the free theory for the environment.
The expectation value of an exponential $e^X$ for a random variable $X$ may be written as the exponential of its cumulants,
\be\label{exp_formula}
\<e^X\> = \exp\[\frac{1}{2}\<X^2\>_c + \frac{1}{4!}\<X^4\>_c+\dots \],
\ee
where the $_c$ subscript indicates the connected part of the correlation function (so $\<X^2\>_c\equiv\<X^2\>-\<X\>^2$), and the dots indicate higher cumulants of $X$.

Considering just a single mode $\zeta_\q$ for our system, and taking the expectation value of the exponent squared, we may write the decoherence rate as the exponential of a loop integral over the environment modes,
\be\label{deco_factor_ImF}
\boxed{ \left.D(\zeta_\q,\tzeta_\q)\right|_\tau = \exp \bigg[ -\frac{4\pi|\Delta\bar{\zeta}_\q|^2}{q^3}\int_{\k+\k'=-\q} P_\E(k,\tau) P_\E(k',\tau) \( {\rm Im}\F_{\k,\k',\q}(\tau) \)^2 + \O(\F^4) \bigg] \, }
\ee
where $\int_{\k+\k'=-\q}\equiv\int\frac{d^3\k}{(2\pi)^3}\frac{d^3\k'}{(2\pi)^3}(2\pi)^3\d^3(\k+\k'+\q)$.
Here we have defined the environment power spectrum,
\be
\<\E_\k\E_{\k'}\>|_\tau \equiv (2\pi)^3 \d^3(\k+\k') P_\E(k,\tau),
\ee
which is related to the variance of the Gaussian wave function,
\be
P_\E(k,\tau) = \frac{1}{2{\rm Re}A_\E(k,\tau)}.
\ee
We have also extracted the $q$-dependence of $|\Delta\zeta_\q|^2$ by defining a rescaled $q$-independent quantity $\bar{\zeta}_\q$, given by\footnote{The factor of $V\equiv(2\pi/\kmin)^{1/3}$ in $\zeta_\q$ cancels with a factor of $\frac{1}{V}$ associated with removing the integral over $\q$ when taking the system to be a single mode.}
\be\label{zeta-bar}
\zeta_\q\equiv\frac{1}{q^{3/2}}V^{1/2}\pi\sqrt{2}\bar{\zeta}_\q,
\ee
with variance equal to the amplitude of fluctuations $\Delta_\zeta^2$,
\be
\<|\bar{\zeta}_\q|^2\>=\frac{H^2}{2\eps\Mp^2}\frac{1}{(2\pi)^2}\equiv\Delta_\zeta^2 \ .
\ee
At early times when the phase oscillations from ${\rm Im}\F$ are small due to the weak coupling, the exponent in Eq. \eqref{deco_factor_ImF} is small and $D(\zeta_\q,\tzeta_\q)\approx1$. On the other hand, when the phase oscillations become large and the exponent (including higher order terms) becomes large, off-diagonal elements in the reduced density matrix become exponentially suppressed.

The decoherence \textit{rate} for a given mode $q$ at time $\tau$ is defined as as minus the exponent of $D(\Delta\zeta_\q)$, evaluated at the average value $\<|\Delta\bar{\zeta}_\q|^2\>=\<|\bar{\zeta}_\q|^2\>+\<|\bar{\tzeta}_\q|^2\>=2\Delta_\zeta^2$,
\be\label{Gamma_def}
\Gamma_{\rm deco}(q,\tau) \equiv -\ln \left.D \(|\Delta\bar{\zeta}_\q|^2=\<|\Delta\bar{\zeta}_\q|^2\>\) \right|_\tau.
\ee
This gives us
\be\label{deco_rate}
\Gamma_{\rm deco}(q,\tau) = \frac{8\pi\Delta_\zeta^2}{q^3}\int_{\k+\k'=-\q} P_\E(k,\tau) P_\E(k',\tau) \({\rm Im}\F_{\k,\k',\q}(\tau)\)^2 + \dots,
\ee
where the dots include higher order terms in the interaction.
When the growth of ${\rm Im}\F$ leads to $\Gamma_{\rm deco}=\O(1)$, the off-diagonal components are becoming exponentially suppressed, indicating that decoherence is occurring.

\begin{figure}
\begin{center}
\includegraphics[scale=0.35]{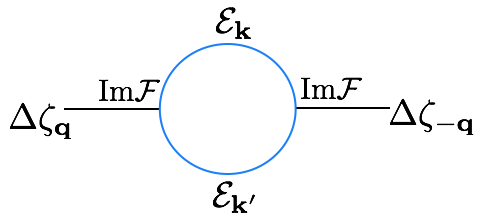}
\end{center}
\caption{The exponent of Eq. \eqref{deco_factor_ImF} grows to $\O(1)$ when decoherence occurs, exponentially suppressing off-diagonal elements ($\zeta_\q-\tzeta_\q\equiv\Delta\zeta_\q\neq0$) of the reduced density matrix $\rho_{\rm R}(\zeta_\q,\tzeta_\q)$. It  is depicted above as a loop integral over the environmental modes which are being traced out (Figure \ref{fig:sum}). The vertex factor comes from the mode coupling in the phase of the wave functional, which is sourced by the cubic interactions in the theory.}
\label{fig:loop}
\end{figure}

As illustrated in Figure \ref{fig:loop}, the decoherence rate is a loop integral, with environment modes running in the loop (but note that the power spectra in the loop are evaluated at the final time $\tau$), with the vertex factor fixed by the interaction Hamiltonian at late times (Eq. \eqref{s_def}).\footnote{Note that although the decoherence rate is $\O(H_{\rm int}^2)$ we do not need to include the quartic part of the wave functional in its computation. When we calculate the exponent on the RHS of Eq. \eqref{exp_formula}, we square the exponent in Eq. \eqref{D_exp_value}, so only the cubic part contributes at leading order. Furthermore, the loop integral determining the decoherence rate should not be affected by renormalization, since quadratic counterterms will not introduce mode coupling and thus will not affect the trace over the environment (although they may affect loop corrections to power spectra).}
However, in this case there is no classical or tree level quantity being corrected. 
As in the case of loop corrections to correlation functions, there is the issues of divergences; when we consider the case of gravitational couplings in \S \ref{sec:zzz}, we will see that the integral is UV finite.
Note also that the decoherence rate scales as $\hbar$; restoring the $\hbar$ dependence,
\be
\Gamma_{\rm deco}\sim\frac{\L_{\rm int}^2}{\hbar^2}\sim\frac{1}{\hbar^2}\(\hbar^{3/2}\)^2 \sim \hbar.
\ee
This is the same dependence as for the amplitude of perturbations, $\<\zeta^2\>\propto\hbar$, which is a quantum mechanical effect (and as a result of which $
\L_{\rm int}\sim\zeta\zeta\zeta\sim\hbar^{3/2}$). In the $\hbar\rarr0$ limit the amplitude of cosmological perturbations vanishes, and decoherence is suppressed.

Anticipating the case of gravitational interactions, we note that Eq. \eqref{deco_factor_ImF} will take the form 
\ba\label{deco_factor_Hint}
&& \hspace{-0.3cm} \left.D(\zeta_\q,\tzeta_\q)\right|_\tau \approx \exp \[ -\frac{4\pi|\Delta\bar{\zeta}_\q|^2}{q^3}\int^{1/|\tau|}_{\k+\k'=-\q} P_\E(k,\tau) P_\E(k',\tau) \int_{-\infty}^\tau d\tau'd\tau'' \tHint_{\k,\k',\q}(\tau')\tHint_{\k,\k',\q}(\tau'') \] , \nn \\
&& \text{for } \tHint\propto a^{p>0}, \ \text{at late times,} \ |q\tau|\ll 1 .
\ea
Here we have used Eq. \eqref{ImF_late} to write the integral in terms of the interaction Hamiltonian.
Note that the momentum integration is restricted to $k,k'<1/|\tau|$. We will see that for the exact expression, Eq. \eqref{deco_factor_ImF}, the integrand drops to zero quickly in the subhorizon regime, justifying the approximation of Eq. \eqref{deco_factor_Hint}.
Effectively, the exponential in Eq. \eqref{F_solution} acts as a theta function $\Theta(1-|k\tau|)$ which turns on after all the modes have crossed the horizon (see Appendix \ref{app:WF}).
As is clear from Eq. \eqref{deco_factor_Hint}, then, decoherence is a superhorizon effect.

Eq. \eqref{deco_factor_ImF}, or equivalently Eq. \eqref{deco_rate}, is the main result for this section, and is approximately given by Eq. \eqref{deco_factor_Hint}. 
If the interaction $H_{\rm int}$ sources a growing phase in the nonlinear part of the wave functional, there are rapid oscillations in the integral over the environment at the epoch when $\Gamma_{\rm deco}\sim1$, suppressing off-diagonal components of the reduced density matrix.
Putting together Eqs. \eqref{deco_factor_ImF} and \eqref{F_solution} (along with Eqs. \eqref{s_def}, \eqref{alpha_def}, and \eqref{Hkin}) gives us a recipe for computing the decoherence rate from the Lagrangian or Hamiltonian, and the free theory evolution $\Psi_G(\tau)$.

\begin{figure}
\begin{center}
\includegraphics[scale=0.45]{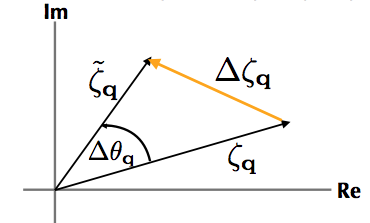}
\hspace{0.5cm}
\includegraphics[scale=0.45]{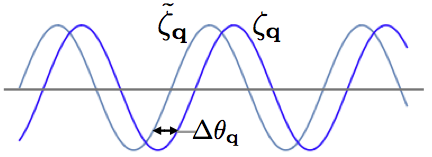}
\end{center}
\caption{As indicated on the right in the complex plane for $\zeta_\q$ and $\tzeta_\q$, two configurations which have either a phase difference (as shown on the left) or amplitude difference will decohere, since $\Gamma_{\rm deco}\propto|\Delta\zeta_\q|^2$.}
\label{fig:amp_phase}
\end{figure}

Before moving on we make a few comments:
\begin{itemize}
\vspace{-0.2cm}
\item
The onset of decoherence is generically delayed due to the weak coupling. The amplitude of the interacting Hamiltonian compared to the free Hamiltonian scales as $\sim\zeta\tHint$, and the decoherence rate scales quadratically with this interaction strength. The amplitude of fluctuations $\Delta_\zeta^2\sim10^{-5}$ contributes part of this suppression. In the case of gravitational couplings of $\zeta$ (see \S \ref{sec:zzz}), the interaction is also slow-roll suppressed, $\tHint\sim\eps$.
\vspace{-0.2cm}
\item
Assuming the integrand only depends on time or scale via the physical wavenumber $-k\tau$ (which we will see in \S \ref{sec:zzz}), 
then upon exchanging the measure $d^3\k$ over comoving wavenumbers for integration over physical scales, the leading time dependence of $\Gamma_{\rm deco}$ goes into an overall scaling with the volume $\sim a^3$, simply as a consequence of the dimensionality \cite{Burgess:2014eoa}. Decoherence will occur when the mode reaches a critical physical wavelength $\lambda_{\rm classical}$ at which this growth overcomes the weak coupling.
\vspace{-0.2cm}
\item
The loop integral adds up the decohering effects of all environment modes. The momentum dependence or ``shape'' of the interacting Hamiltonian kernel, $\tHint_{\k,\k',\q}$, determines the physical wavelength of the environment modes which have the strongest decohering effect. The environment grows as more modes redshift to this scale.
\vspace{-0.2cm}
 \item Either an amplitude or phase difference between $\zeta_\q$ and $\tzeta_\q$ can lead to decoherence. The amplitude and phase of a mode are defined by $\zeta_\q\equiv|\zeta_\q|e^{i\theta_\q}$, and the difference of two complex numbers in terms of the modulus and phase differences is $|\Delta\zeta_\q|^2=\Delta|\zeta_\q|^2+2|\zeta_\q||\tzeta_\q|(1-\cos(\Delta\theta_\q))$. Both amplitude and phase differences contribute to $|\Delta\zeta_\q|$, so either is sufficient for the two configurations to decohere (see Figure~\ref{fig:amp_phase}).\footnote{Note that the dependence on $|\Delta\zeta_\q|^2$ is enforced by translation invariance, which ensures that only momentum-conserving triplets of wavenumbers are coupled.}
\end{itemize}

\section{Decoherence from Gravitational Interactions}
\label{sec:zzz}

We now specify to the case where fluctuations of $\zeta$ itself act as the environment, and the system-environment coupling results from minimal gravitational interactions \cite{Maldacena:2002vr}. We will set $c_s=1$ and comment on the more general case in Appendix \ref{app:cs}.

From the quadratic Lagrangian for $\zeta$, Eq. \eqref{S2zeta}, the conjugate momentum $\pi=\del\L/\del\zeta$ is\footnote{When computing the Hamiltonian density $\H=\pi\dot{\zeta}-\L$, the quadratic corrections to $\pi$ generate cubic terms in both $\pi\dot{\zeta}$ and from the free Lagrangian, but these cancel, so that we can effectively drop the corrections in Eq. \eqref{pi_zeta}, and set $\H_{\rm int}=-\L_{\rm int}$ up to cubic order \cite{Huang:2006eha}.}
\be\label{pi_zeta}
\pi_\k
= 2\eps\Mp^2 a^3\dot{\zeta}^*_\k + \O\big((\zeta,\dot{\zeta})^2\big),
\ee
and so the free Hamiltonian is
\be
H_\zeta = \frac{1}{2}\int\frac{d^3\k}{(2\pi)^3}
\(\frac{1}{2\eps\Mp^2}\frac{1}{a^3} \pi_\k\pi_{-\k}
+ 2\eps\Mp^2 a k^2 \zeta_\k\zeta_{-\k}\).
\ee
Consequently, from Eq. \eqref{Hkin} we have
\be
f_\zeta(\tau) = -\frac{H^3\tau^3}{2\eps\Mp^2}. \label{f_zeta}
\ee

In the case that $\zeta$ starts in the Bunch-Davies vacuum in the $t_0\rarr-\infty$ limit, the Schr\"{o}dinger picture evolution of the state is \cite{Polarski:1995jg,Burgess:2014eoa}
\be\label{AzetaBD}
A_\zeta(k,\tau) = 2k^3 \frac{\eps\Mp^2}{H^2} \frac{1-\frac{i}{k\tau}}{1+k^2\tau^2}.
\ee
While the real part becomes constant in the late-time limit, reflecting the freezing of $\zeta$ in the superhorizon regime, the imaginary part grows with the scale factor, contributing to the squeezing of the quantum state. (Specifically, the growing phase describes the ratio of squeezing to freezing. One can show that $\sqrt{\<|\dot{\zeta}_\q|^2\>}\sim\theta(a)/a^3\sim1/a^2$, where the phase $\theta(a)$ is the imaginary part of $A_\zeta$. The phase oscillations slow down the freezing of $\zeta$ compared to the squeezing of the state, which becomes peaked in phase space for ``classical'' values $\dot{\zeta}_{\rm cl}=\dot{\zeta}(\zeta)$ \cite{Kiefer:2008ku} more rapidly than $\dot{\zeta}\rarr0$. This behavior is a consequence of the growth of the action, which leads to phase oscillations, $\L\propto a(t)$.)

The most universal couplings we can consider are purely gravitational self-interactions of $\zeta$ (or interactions with tensor modes $\gamma_{ij}$), due to the nonlinearity of GR \cite{Maldacena:2002vr}. We will work in the gauge where matter fluctuations vanish on equal-time hypersurfaces\footnote{Of course, it may not be possible to gauge away matter fluctuations if multiple field are present, but in this case additional couplings could only amplify the minimal decoherence rate which we will find for gravitational couplings.}, leaving only metric fluctuations, and will address the issue of gauge dependence in \S \ref{sec:cons_rel}.
Self-interactions of $\zeta$ have a ``coupling constant'' $\eps$, and include terms proportional to
\be\label{zzzterms}
a^3\zeta\dot{\zeta}^2, \ \ a\zeta(\del_i\zeta)^2), \ \ a^3\zeta^2\dot{\zeta}, \ \ a^3\dot{\zeta}(\del_i\zeta)(\del_i(\del^{-2}\dot{\zeta})), \ \ \text{etc.}
\ee
The time dependence of these interactions, controlled by the scale factors and factors of $\dot{\zeta}$, determines their relevance for the late-time behavior of $\tHint$, which determines ${\rm Im}\F$.
Most terms do not lead to a growing source, $\tHint \sim a(t)$ at late times, because $\dot{\zeta}\rarr0$.\footnote{\label{footnote_zzz_terms}The cubic Lagrangian for $\zeta$ is given in Eq. $(3.9)$ in \cite{Maldacena:2002vr}, or more generally Eq. $(4.26)$ in \cite{Chen:2006nt}, including a nontrivial speed of sound. Converting from $\dot{\zeta}$ to the conjugate momentum, Eq. \eqref{pi_zeta}, introduces a factor of $1/a^3$, so terms with $\dot{\zeta}$ are more suppressed in the late-time limit.
(Note also that further acting $\pi=-i\d/\d\zeta$ on $\Psi_G$ to obtain the source in Eq. \eqref{s_def} contributes an imaginary factor ${\rm Im} A_\zeta\sim a$ to the source $\tHint$, while ${\rm Re} A_\zeta\rarr{\rm const}$; this does not overcome the $1/a^3$ suppression of $\dot{\zeta}$.)
In addition to the terms shown in Eq. \eqref{L_zzz}, there is another term, $\frac{1}{2}\eps\dot{\eta}a^3\zeta^2\dot{\zeta}$,
which produces a growing source for ${\rm Im}\F$ in the late time limit. However, this is higher order in slow roll parameters, so we ignore it.}
It is a consequence of this freezing behavior that the relevant late-time interactions involve only the field and not its conjugate momentum, and thus commute with $\hat{\zeta}$, indicating that decoherence takes place in the field basis \cite{Kiefer:2006je,Burgess:2014eoa}.
We focus on the two terms which do not freeze to zero, and which therefore may source a rapidly oscillating phase and lead to decoherence, which are \cite{Maldacena:2002vr}
\ba
\L_{\zeta\zeta\zeta} &=& \Mp^2 \( \eps^2 a\zeta(\del\zeta)^2 - \frac{1}{2} \eps\eta a \zeta^2\del^2\zeta \) + ... \nn \\
&=& -\frac{\Mp^2}{2}\eps(\eps+\eta)a\zeta^2\del^2\zeta+..., \label{L_zzz}
\ea
where we have integrated by parts in the second line.\footnote{The second term in the first line of Eq. \eqref{L_zzz} is among a large collection of terms that may be removed by a field redefinition, which is useful for computing correlation functions \cite{Maldacena:2002vr,Chen:2006nt}. We include it here, because we are interested in the reduced density matrix of $\zeta$ rather than for a redefined field.}
Our interaction Hamiltonian is then
\be
H_{\rm int} = \frac{\Mp^2}{2} \int d^3\x\eps(\eps+\eta) a\zeta^2\del^2\zeta,
\ee
so from Eq. \eqref{s_def} we have
\be
\tHint_{\k,\k',\q}(\tau) = \frac{\eps\Mp^2}{6H\tau} (\eps+\eta)(k^2+k'^2+q^2). \label{source_zzz}
\ee

In Appendix \ref{app:WF} we use Eqs. \eqref{f_zeta}-\eqref{AzetaBD} and Eq. \eqref{source_zzz} to compute $\F_{\k,\k',\q}(\tau)$ as given in Eq. \eqref{F_solution}.
This calculation shows that the integrand in Eq. \eqref{F_solution} ``turns on'' when all the modes cross the horizon; that is, $\exp\[i\int_{\tau'} d\tau''\alpha_{k,k',q}(\tau'')\]\sim\Theta(1-|k\tau'|)$.
We will be interested in the case where the system mode is deeply superhorizon, $|q\tau|\ll1$. At the same time, in order to integrate over the environment modes, we would like to know ${\rm Im}\F$ when the environment modes can still be in the horizon crossing regime.
The solution is given in Eq. \eqref{ImF_zzz_qeta},
\ba
{\rm Im}\F_{\k,\k',\q}(\tau) &=& - \frac{\eps\Mp^2}{6H^2}\frac{1}{\tau}(\eps+\eta)\frac{k^2+k'^2+q^2}{(1+k^2\tau^2)(1+k'^2\tau^2)} + \O(q\tau) \hspace{1cm} (q\ll1/|\tau|) \nn \\
&\rarr&  -\frac{\eps\Mp^2}{3H^2}(\eps+\eta)\frac{1}{k\tau}\frac{k^3}{(1+k^2\tau^2)^2}.  \hspace{1cm} (q\ll k,k',|\tau|^{-1}) \label{ImF_zzz}
\ea
In the second line we have taken the squeezed limit $q\ll k,k'$.
Note that while the phase ${\rm Im}\F$ grows as $1/\tau\sim a(t)$, the real part of $\F$ -- which we have omitted here -- remains small so that $|\Psi|^2_{\tau\rarr0}$ is nearly Gaussian, and determines the $\zeta(\del\zeta)^2$ contribution to the three-point function $\<\zeta_{\k}\zeta_{\k'}\zeta_{\q}\>$ (see Appendix \ref{app:WF}).

\begin{figure}
\begin{center}
\includegraphics[scale=0.5]{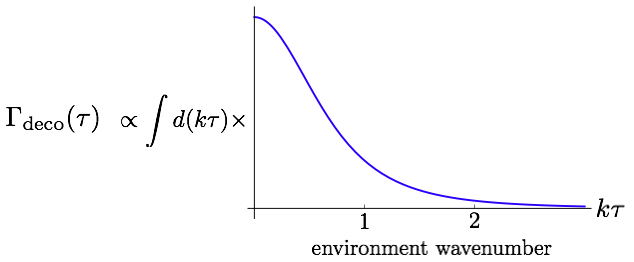}
\end{center}
\caption{The function of environment wavenumber $k$ which is integrated over to obtain the decoherence rate $\Gamma_{\rm deco}$. The integrand is $(k\tau)^2({\rm Im}\F_{k,k',q})^2/A_\zeta(k,\tau)^2\sim(1+k^2\tau^2)^{-2}$. The integral mainly receives contributions from modes that have just crossed the Hubble scale, $|k\tau|=\O(\frac{1}{2})$ or so, and only receives a small contribution from very superhorizon modes, $|k\tau|\ll1$.}
\label{fig:integrand}
\end{figure}

We now consider a single mode $\zeta_\q$, treating other modes as the environment.
The decoherence factor $D(\zeta_\q,\tzeta_\q)$ as given in Eq. \eqref{deco_factor_ImF} depends on the environment power spectrum,
\be\label{ReA}
P_\E(k,\tau) = P_\zeta(k,\tau) = \frac{H^2}{2\eps \Mp^2} \frac{1}{2k^3}(1+k^2\tau^2).
\ee
Plugging Eqs. \eqref{ImF_zzz} and \eqref{ReA} into Eq. \eqref{deco_factor_ImF} and simplifying the integral, which is dominated by squeezed configurations $k,k'\gg q$, we find
\be\label{d_zzz}
\left.D(\zeta_\q,\tzeta_\q)\right|_\tau = \exp\[ - \frac{(\eps+\eta)^2}{72\pi}|\Delta\bar{\zeta}_\q |^2\frac{1}{|q\tau|^3}\int \frac{d(k\tau)}{(1+k^2\tau^2)^2} + \O(\F^4) \],
\ee
where $|\Delta\bar{\zeta}_\q|^2\equiv (q^3/2\pi^2 V) |\Delta\zeta_\q|^2=O(\Delta_\zeta^2)$ was defined in Eq. \eqref{zeta-bar} as the rescaled dimensionless amplitude of $\zeta_\q - \tzeta_\q$.
We see that the only dependence on time or on the short mode $k$ is in the physical wavenumber $-k\tau=k/aH$, leading to an overall $\sim a^3$ growth as noted in \S \ref{sec:dec_factor}.
At small $k$ the integrand in Eq. \eqref{d_zzz} grows linearly in $k$ with respect to a logarithmic measure $d\ln k$, before falling off for $|k\tau|\gtrsim1$, so the integral is dominated by modes that have just crossed the horizon -- see Figure \ref{fig:integrand}. We emphasize that the integral is convergent in the UV, unlike loop integrals in correlation functions involving the real part of $\F$, which need to be renormalized.
Evaluating the integral and setting $\tau=-1/aH$, we finally have
\ba\label{d_boxed_final}
&&\boxed{\left.D(\zeta_\q,\tzeta_\q)\right|_{a} =
\exp\[-\frac{1}{288}(\eps+\eta)^2|\Delta\bar{\zeta}_\q|^2\(\frac{aH}{q}\)^3 + \dots \]. } \hspace{0.7cm} (q\ll aH)
\ea
The dots indicate additional terms which also become large when the leading $\O(\eps^2|\Delta\zeta|^2)$ term becomes large. Equivalently, the decoherence rate is
\be\label{Gamma_zzz}
\Gamma_{\rm deco}(q,a) \approx \(\frac{\eps+\eta}{12}\)^2\Delta_\zeta^2\(\frac{aH}{q}\)^3. \hspace{0.7cm} (q\ll aH)
\ee

\begin{figure}
\begin{center}
\includegraphics[scale=0.75]{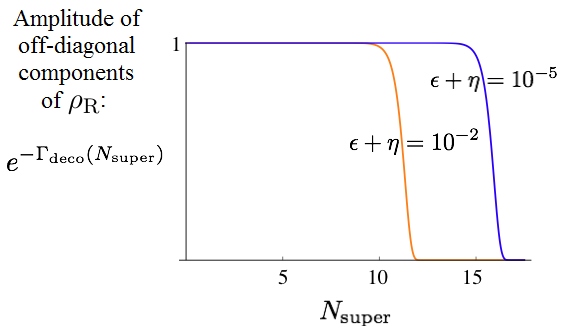}
\end{center}
\caption{The decoherence factor $D=\exp[-\Gamma_{\rm deco}]$ for a given mode, in terms of the number of $e$-folds after horizon crossing, evaluated for a typical off-diagonal component of the reduced density matrix, $|\Delta\zeta_\q^2|=\<|\Delta\zeta_\q^2|\>$, and fixing $\Delta_\zeta^2=2.5\times 10^{-9}$. We see that off-diagonal components are rapidly suppressed after $10$-$20$ $e$-folds, with a decay time of order the Hubble time, $\O(1)$ $e$-fold.}
\label{fig:D(N)}
\end{figure}

We see that the decoherence rate grows as the number of physical Hubble volumes at time $\tau=-1/aH$, as described earlier.
However, it is suppressed by $\eps^2|\Delta\zeta_\q|^2$ due to the smallness of the interaction.
When the environment modes cross the horizon, the phases from couplings to the superhorizon system mode grow $\sim a(t)$.
\textit{When $\Gamma_{\rm deco}=\O(1)$, this results in many phase oscillations in the integral over the environment, Eq. \eqref{rho_env_int}, quickly suppressing off-diagonal elements of $\rho_{\rm R}(\zeta_\q,\tzeta_\q)$.}
Setting $\Gamma_{\rm deco}=1$ we see that, as illustrated earlier in Figure \ref{fig:QChorizon}, there is a delay of 
\be
N_{\rm deco} \approx \frac{2}{3} \ln \(\frac{100}{\eps+\eta}\frac{\sqrt{\eps}\Mp}{H} \)
\ee
$e$-folds after a mode crosses the horizon but before it decoheres.
In Figure \ref{fig:D(N)} we plot the decoherence factor as a function of $e$-foldings after horizon crossing, for two values of $\eps+\eta$.
Because the mode being decohered is very superhorizon, while the environment modes are close to the horizon scale, it is indeed squeezed configurations $k,k'\sim aH\gg q$ that are predominant in decohering $\zeta_\q$.

\subsection{Super-Hubble Environment Modes}
\label{sec:IR}

\begin{figure}
\begin{center}
\includegraphics[scale=0.35]{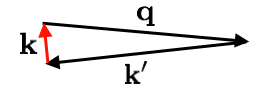}
\end{center}
\vspace{-0.5cm}
\caption{Super-Hubble modes $k\lesssim q$ which have already decohered can act as an environment, but are a subdominant source of decoherence.}
\label{fig:IRtriangle}
\end{figure}

In the previous section, we only showed the dominant contribution from Hubble-scale environment modes, obtained from the limit of the second line of Eq. \eqref{ImF_zzz}.
Using the exact expression and including the contribution from super-Hubble environment modes $k\lesssim q$, we find a subleading contribution to the decoherence rate,\footnote{The dominant term $\Gamma^{(k\sim aH)}_{\rm deco}$ came from taking the Laplacian in the interaction $\zeta^2\del^2\zeta$ to act on (short-wavelength) environment modes, that is $\del^2\rarr k^2$. If we instead consider the term with the derivatives on the system mode, $\del^2\rarr q^2$, then we find the subdominant but IR divergent term in Eq. \eqref{Gamma_IR}.}
\be\label{Gamma_IR}
\Gamma_{\rm deco} = \Gamma^{(k\sim aH)}_{\rm deco} + \frac{(\eps+\eta)^2\Delta_\zeta^2}{9\pi}\(\frac{aH}{q}\)^2 \[\ln\(\frac{q}{\kmin}\) -\frac{19}{48} \] + \O(\eps,\eta)^4,
\ee
where $\Gamma^{(k\sim aH)}_{\rm deco}$ is the dominant contribution, Eq. \eqref{Gamma_zzz}. 
The second term comes from superhorizon environment modes $k\lesssim q$ (see Figure \ref{fig:IRtriangle}), including both ``equilateral'' configurations with $k,k'\sim q$, and a log-enhanced contribution from very soft momenta $k\ll q$ running in the loop integral. It grows more slowly, as the area instead of the volume, but diverges in the IR as $\ln\kmin$.\footnote{\label{log_scaledep}We have assumed a scale invariant spectrum. In reality, when we integrate deep into the IR, even a small spectral index $n_s=\O(\eps,\eta)$ will lead to a large difference.  More generally, the factor of $\Delta_\zeta^2\ln(q/\kmin)$ in Eq. \eqref{Gamma_IR} will be replaced with a factor $\sim\<\zeta_L^2\>=\int^q_{\kmin}\frac{d^3\k}{(2\pi)^3}P_\zeta(k)$, where $\zeta_L$ includes long-wavelength modes $k\lesssim q$, and the power spectrum is slightly scale-dependent.}
Considering the time when decoherence occurs, fixed by $\Gamma^{(k\sim aH)}_{\rm deco}\sim1$ or equivalently $q/aH\sim(\eps H^2/\Mp^2)^{1/3}$, we see that the subleading term is larger and appears to be the leading source of decoherence after a very large number of $e$-folds,
\be
N_{\rm IR}\sim\bigg(\eps\frac{H^2}{\Mp^2}\bigg)^{-1/3},
\ee
have passed (here we assume $\eps+\eta=\O(\eps)$ for simplicity). At this point, there has accumulated a large super-Hubble environment of modes which have themselves already decohered. However, in this regime we do not fully trust our calculation. The Hubble scale changes by
\be
\frac{\Delta H}{H}\sim\eps N_{\rm IR} \sim \bigg(\frac{\eps\Mp}{H}\bigg)^{2/3},
\ee
while the overall amplitude of fluctuations is\footnote{When we include scale-dependence in the power spectrum, we no longer have $\<\zeta^2\>\sim\Delta_\zeta^2 N_{\rm IR}$. However, because $\<\zeta^2\>$ is the quantity that appears in Eq. \eqref{Gamma_IR} (see footnote \ref{log_scaledep}), it is still true that $\<\zeta^2\>\sim(\eps\Mp/H)^{-4/3}$.}
\be\label{zeta_dH}
\<\zeta^2\>\sim\Delta_\zeta^2 N_{\rm IR}\sim \bigg(\frac{\eps\Mp}{H}\bigg)^{-4/3} \sim\bigg(\frac{\Delta H}{H}\bigg)^{-2}.
\ee
So when the length of inflation exceeds $N_{\rm IR}$, either the Hubble constant has changed by an $\O(1)$ factor, or the fluctuations are no longer perturbatively small. In either case our assumption of a quasi de Sitter background with small fluctuations of matter and geometry fails. Consequently, in a homogeneous nearly de Sitter background, the IR environment cannot be large enough to compete with the Hubble-scale UV environment.

While modes which have already decohered or are decohering at the same time as the system mode are a subdominant source of decoherence, the $\sim a^2$ growth in Eq. \eqref{Gamma_IR} shows that they are responding to the system mode. Much longer modes $q\ll k$ which have already decohered may respond stochastically to the system mode as classical fluctuations. 

\subsection{Squeezed-Limit Consistency Relations and Gauge Dependence}
\label{sec:cons_rel}

The dominant role of the squeezed limit in decoherence may raise eyebrows, in light of consistency relations in single-clock inflation which tightly constrain the squeezed limits of inflationary correlation functions.
The gravitational couplings that produced decoherence also contribute to the three-point function $\<\zeta_\k\zeta_{\k'}\zeta_{\q}\>$ or bispectrum, which is determined by ${\rm Re}\F_{\k,\k',\q}$.
In single-clock inflation, the squeezed-limit of the three-point function is fully determined by the two-point function \cite{Maldacena:2002vr,Creminelli:2004yq,Cheung:2007sv},
\be\label{cons_rel}
\<\zeta_\q\zeta_\k\zeta_{\k'}\>' = P(q)P(k)\[-(n_s-1)+ \O(q^2/k^2) \],
\ee
where the prime indicates removal of $(2\pi)^3\d^2(\k+\k'+\q)$, the power spectrum is defined by $\<\zeta_\k\zeta_{\k'}\>\equiv(2\pi)^3\d^3(\k+\k')P(k)$, and $n_s-1\equiv d\ln P(k)/d\ln k$.
Upon making a coordinate transformation from global comoving coordinates to physical coordinates appropriate for a local observer,
\be
k_{\rm ph}(k) = k e^{-\zeta_L} \approx k(1-\zeta_L),
\ee
where the long-wavelength background $\zeta_L\simeq\int^k\frac{d^3\q}{(2\pi)^3}\zeta_\q$ rescales the small-scale coordinates,
this correlation vanishes in the $q\rarr0$ limit\footnote{The RHS of Eq. \eqref{phys_sq_lim} scales as $1/q$ because $P(k)\propto1/q^3$, but the appropriate rescaled correlation function $\<\zeta_\q\zeta_\k\zeta_{\k'}\>/\sqrt{P(q)P(k)P(k')}$ vanishes as $q^{1/2}$.}: a long-wavelength background has no effect on short-wavelength fluctuations \cite{Pajer:2013ana,Tanaka:2011aj},
\be\label{phys_sq_lim}
\<\zeta_\q\zeta_{\k_{\rm ph}(\k)}\zeta_{\k'_{\rm ph}(\k')}\>' = P(q)P(k)\times\O(q^2/k^2).
\ee
The apparent squeezed-limit correlation in Eq. \eqref{cons_rel} vanishes in the physical short-scale coordinates defined by the long-wavelength mode.

This raises the question of whether the squeezed-limit coupling which produced decoherence, and arose from interactions which satisfy the above consistency relation, is indeed a physical effect. We will not attempt to fully resolve this issue here, but will offer a few comments.

First, while squeezed-limit configurations appear to be the dominant source of decoherence, ``equilateral'' configurations in which $k_{\rm environment}=\O(q)$ contribute at a subdominant level. 
As noted in \S \ref{sec:IR}, including equilateral configurations gives a contribution
\be
\Gamma^{(k\ll aH)}_{\rm deco} \sim \eps^2|\Delta\zeta|^2(aH/q)^2,
\ee
scaling as the area instead of the volume (this is simply a consequence of the $(\L_{\zeta\zeta\zeta})^2\sim a^2$ dependence). So in the absence of squeezed-limit couplings, decoherence would be delayed several $e$-folds until $\Gamma^{(k\ll aH)}=\O(1)$, but is inevitable due to the growing phase oscillations in the wave functional, which are present for all configurations $(k,'k',q)$. Thus, decoherence does not rely entirely on the squeezed-limit coupling.

Second, the interaction which produces decoherence scales differently than the squeezed-limit three-point function amplitude, $n_s-1=-2\eps-\eta$, which is removed by a coordinate transformation. While decoherence relies only on the $a\zeta(\del\zeta)^2$ interaction, the squeezed bispectrum also receives a contribution from the $a^3\zeta\dot{\zeta}^2$ interaction \cite{Chen:2006nt}. This term does not produce a growing phase in $\Psi[\zeta]$ and hence decoherence, essentially because $\dot{\zeta}^2\sim1/a^4$ vanishes faster than $(\del\zeta)^2/a^2$ in the late-time limit. So the same coordinate transformation that removes the squeezed-limit three-point function would not remove the interaction which produces decoherence.

Moreover, while we have calculated the decoherence rate in a specific gauge (the time slicing in which matter fluctuations vanish on constant time hypersurfaces), we could certainly have worked in a different gauge, such as matter fluctuations on slices of constant spatial curvature, or worked with the Goldstone mode $\pi$ from breaking of time translations. The interaction which produces decoherence is not an artifact of the gauge we have chosen.

It is true that the $\zeta(\del\zeta)^2$ interaction could be removed by a change of variables to $\zeta_{\rm NL}=\zeta+\frac{1}{2}(\eps+\eta)\zeta^2$. However, this is not a field redefinition that would rescale short-scale quantities in accordance with a long-wavelength background.  
It is always possible to choose a particular variable $\zeta_{\rm NL}$ for which the interactions generating decoherence (including couplings to tensor modes; see \S \ref{sec:disc}) are removed, but we are not aware of a reason why that particular variable would be physically relevant.
Choosing a particular system-environment decomposition for which off-diagonal components of the reduced density matrix are not suppressed does not mean that there is no decoherence; rather, the decomposition chosen does not show the decoherence.

Lastly, we are studying decoherence of Fourier modes in a global inflating volume, and while superhorizon modes are long-wavelength, they are not a constant background or constant gradient for the Hubble-scale environment. (Of course, it would be interesting to decompose field configurations of $\zeta$ into basis functions that are spatially localized, as sketched in \S \ref{sec:future}, so that we could see explicitly the effect of coupling to a long-wavelength background.) As noted in \S \ref{sec:IR}, our calculation is only valid for superhorizon modes that are not too long-wavelength, and fails in the $q\rarr0$ limit. The effect we have described arises from finitely squeezed configurations, with a small but finite ratio $k_{\rm long}/k_{\rm short}$.


If the squeezed limit is indeed the dominant source of decoherence, then the picture that emerges is one where Hubble-scale modes act as a measuring device for the long-wavelength background $\zeta_L(\x)\sim\int d^3\q \zeta_\q e^{i\q\cdot\x}$ for a region around point $\x$. This background sets the local amount of expansion in any given region, which acts as a local time coordinate, so short modes act as a ``clock'' to record the local time.
The sensitivity of short modes in this sense is due to the growing phase of the wave functional, and is different from the (in)sensitivity of short-wavelength correlation functions to a long-wavelength background, which is instead determined by the probability $|\Psi|^2$.

\section{Discussion}
\label{sec:disc}

\subsection{Summary}

Our results, overviewed earlier in \S \ref{sec:results} and Figure \ref{fig:QChorizon}, suggest that gravitational nonlinearities from GR are sufficient to generate classical stochastic perturbations during inflation from quantum fluctuations stretched to very superhorizon scales.
As each mode redshifts to the Hubble scale, it becomes sensitive as an environment for longer modes, $k_{\rm env}\sim aH$. After further redshifting for several e-folds and reaching $k_{\rm classical}/a\sim H(\eps H^2/\Mp^2)^{1/3}$, it is decohered by shorter Hubble-scale modes, and evolves from a pure to mixed state with a diagonal density matrix:\footnote{Note that the density matrix, Eq. \eqref{diag_rho}, will depend on the realization of longer-wavelength modes which have already decohered at a very small level, due to the non-Gaussianity from gravitational couplings.}
\be\label{diag_rho}
\rho_{\rm R}(\zeta_\q,\tzeta_\q)\rarr\rho_{\rm R}(\zeta_\q,\zeta_\q)\delta^{(2)}(\zeta_\q-\tzeta_\q).
\ee
This is a consequence of growing phase oscillations in the wave functional due to the inflationary growth in the action, $\L_{\rm int}\propto a(t)$.
Thus, because the wave functional evolves into a WKB state with rapidly oscillating phases, there is also a transition to classicality in the sense of decoherence. As noted in \S \ref{sec:zzz}, the related growth of the free theory action, $\L_{\rm free}\sim a(t)$, underlies the squeezing of the quantum state.

As seen in Figure \ref{fig:D(N)}, the off-diagonal components become exponentially suppressed within $\O(1)$ e-folds. That is, the decoherence timescale is of order the Hubble time. Superhorizon modes $\zeta_\q$ are freezing and evolve on a much longer timescale, $\dot{\zeta}_\q/\zeta_\q\sim(q\tau)^2H\ll H$, and are therefore measured very rapidly compared to their own evolution.
As emphasized in previous works \cite{Burgess:2014eoa,Kiefer:2006je}, the ``pointer basis'' -- the basis of the variable being measured or decohered -- is the field configuration basis.

Squeezed limit mode coupling plays a key role in the decoherence process: As a given mode $q$ redshifts into the classical regime, the environment modes which carry away the largest phases and thus contribute most to decoherence are pairs of short modes which couple to it in the squeezed limit, $k,k'\gg q$.
Physically, the average value of $\zeta$ in an inflating region is the amount of expansion, which acts as a local time variable.  Shorter-wavelength modes of $\zeta$ therefore play the role of recording the time in a given region.  This relies on the breaking of time translation invariance through nonzero slow roll parameters, so that there is a genuine dynamical time variable to be measured.

While the inflating background makes decoherence inevitable, it is delayed by the weakness of the coupling. Introducing additional degrees of freedom with stronger couplings will likely speed up decoherence, but if the interactions are characterized by a small dimensionless coupling strength $g$, then our results suggest that it will occur when the inflating volume overcomes this suppression:
\be
\(\frac{q}{aH}\)^3 \sim g^2,
\ee
or $\sim\frac{1}{3}\ln(g^{-2})$ $e$-folds after horizon crossing.
We emphasize that decoherence occurs entirely in the superhorizon regime: In the subhorizon regime, where modes act as they would in the Minkowski space vacuum, there are no rapidly oscillating phases. This is consistent with the expectation that flat-space vacuum fluctuations should not by themselves decohere \cite{Kiefer:2006je,Burgess:2006jn}. 

We also emphasize that gravitational interactions can turn quantum fluctuations of the metric into classical spacetime perturbations, given an inflating background.
In addition to the scalar self-interactions discussed above, nonlinearities in general relativity couple tensor modes $\gamma$ to scalar modes \cite{Maldacena:2002vr}:
\be\label{L3gz}
\L_{\zeta\zeta\gamma}\sim\eps a\zeta\del_l\gamma_{ij}\del_l\gamma_{ij},
\hspace{1cm}
\L_{\zeta\gamma\gamma}\sim\eps a\gamma_{ij}\del_i\zeta\del_j\zeta.
\ee
These couplings only differ from the $\zeta(\del_i\zeta)^2$ coupling for scalar modes in their tensor structure, and should therefore also lead to decoherence. So we expect that \textit{Hubble-scale graviton modes contribute to the decoherence of scalar curvature perturbations, and Hubble-scale scalar modes lead to decoherence of background gravitational waves.}
It would be interesting to know whether graviton self-interactions \cite{Maldacena:2002vr,Maldacena:2011nz} by themselves lead to a classical gravitational wave background.

The presence of tensor modes makes decoherence even more unavoidable.
With only the scalar mode, tuning $\eps+\eta=0$ -- which corresponds to $H(t)\propto1/\sqrt{2t+\rm const.}$ --
would turn off the relevant system-environment coupling and prevent decoherence.
However, the couplings to tensor modes, which scale as $\eps$, would still be present, suggesting that decoherence would still occur.

Lastly, we note that no initial state excitations above the Bunch-Davies vacuum (which was imposed by fixing Eq. \eqref{AzetaBD} and $\F(t\rarr-\infty)=0$) are needed. A decelerating cosmology, on the other hand, may need particles in order for metric fluctuations to decohere.

In summary, the inflating background allows for a superhorizon regime, which, along with the existence of minimal gravitational couplings, is sufficient to decohere vacuum fluctuations and produce classical stochastic perturbations.

\subsection{Comparison to Previous Work}
\label{sec:lit}

There is of course an extensive body of work on the quantum-to-classical transition during inflation \cite{Kiefer:2008ku,Polarski:1995jg,Albrecht:1992kf,Grishchuk:1989ss,Grishchuk:1990bj,Guth:1985ya,Kiefer:1998qe,Lombardo:2005iz,Calzetta:1995ys,Franco:2011fg,Sakagami:1987mp,Brandenberger:1990bx,Prokopec:2006fc,Sharman:2007gi,Burgess:2014eoa,Burgess:2006jn,Martineau:2006ki}, and a number of these works have studied decoherence of inflationary perturbations due to an environment.

Our general result for the decoherence rate, Eq. \eqref{deco_rate}, is complementary to Burgess et. al. \cite{Burgess:2014eoa}, who found that a decoherence rate $\Gamma\propto a^3$ arises generically for couplings to an environment of the form $H_{\rm int}=\int d^3\x a^3 \zeta(\x)\mathcal{B}(\x)$, where $\mathcal{B}$ is a functional of environmental field(s) satisfying certain conditions. (In the case of \S \ref{sec:zzz}, $\mathcal{B}\propto(\del\zeta)^2/a^2$.)
In comparison, our approach highlights the role of phase oscillations in the wave functional and the relation of decoherence to WKB classicality, as well as the relation to non-Gaussianity -- especially in the squeezed limit. We also identify the time during inflation at which a given mode decoheres, as well as which environment modes induce decoherence. 

In our calculation we did not assume that the evolution of the reduced density matrix is Markovian or that it could be described with the Lindblad equation,
but have instead evolved the global wave functional. However, we have found that a super-Hubble system mode is decohered by Hubble-scale environment modes with a correlation time much shorter than the timescale associated with the interactions coupling system and environment.
Consequently, we expect the evolution of the reduced density matrix to be approximately Markovian, with the reduced density matrix being approximately governed by the Lindblad equation \cite{Burgess:2014eoa}. At the same time, computing the exact evolution of the wave functional ensured UV convergence for the decoherence rate, whereas the decoherence rate obtained from Lindblad evolution depends on a correlation function of environment modes which may need to be renormalized \cite{Burgess:2014eoa}.

The present work is also similar to \cite{Franco:2011fg}, which showed the slow-roll suppression and exponential enhancement of decoherence for gravitational couplings of scalar and tensor modes. Here we have also found the time of decoherence and dominant environmental modes.
Also related is \cite{Martineau:2006ki}, which studied decoherence from gravitational couplings, although not the minimal couplings of \cite{Maldacena:2002vr}.

One might ask whether gravitational nonlinearities matter for decoherence, if there may be stronger interactions from the matter sector. (Indeed, gravitational nonlinearities are smaller than others, being suppressed both by slow roll parameters and by the amplitude of fluctuations $\Delta_\zeta^2$.) If any other interactions are present, such as inflaton self-interactions \cite{Lombardo:2005iz} or couplings to additional fields \cite{Sakagami:1987mp,Brandenberger:1990bx,Prokopec:2006fc}, they very well may be irrelevant. However, our emphasis is that \textit{even if no other interactions arise from the matter sector, decoherence will still occur and arises necessarily as a consequence of gravity.}
Even in the vanilla case of $V(\phi)=\frac{1}{2}m^2\phi^2$, the interactions studied here are present.
\textit{Gravitational couplings thus put a lower bound on the decoherence rate during inflation.}
We also note that in the effective field theory for single-field inflationary fluctuations \cite{Cheung:2007st}, the leading cubic interactions (from matter self-interactions) are $\dot{\zeta}$-suppressed and do not contribute to decoherence, so (subleading) gravitational interactions may be the dominant source of decoherence for general single-field models beyond slow-roll (see Appendix \ref{app:cs}).
Beyond single-field, even minimal coupling of matter and gravity will introduce many new couplings of $\zeta$ to additional degrees of freedom, which will lead to an earlier time of decoherence

\subsection{Observability, the Quantum Origin of Perturbations, and Stochastic Inflation}
\label{sec:obs_q_grav}

In this section we address the question of whether inflationary decoherence has observational relevance.

In the minimal setting of the scalar mode $\zeta$ with gravitational interactions, we do not expect any observable signature of the decoherence process.
Because the conjugate variable $\dot{\zeta}$ is observationally inaccessible, all expectation values can be reproduced as cumulants of a classical probability distribution $\rho[\zeta(\x)]$.
The off-diagonal components of the reduced density matrix cannot be traced observationally even in the absence of any nonlinearities or decoherence.

While post-inflationary observables will be governed by a classical PDF, it is still possible for those classical statistics to reveal the quantum origin of perturbations in scenarios beyond single-clock inflation. The possibility of violations of Bell's inequalities in a cosmological setting was recently discussed in \cite{Maldacena:2015bha}; here, the analogue of a Bell experiment takes place during inflation, with the outcome of the experiment being recorded in the resulting classical statistics of perturbations.
In laboratory-based experiments testing Bell inequalities, entangled EPR pairs of particles must remain isolated from environmental degrees of freedom until measured, or else the coherent superposition of states of the entangled pair will be destroyed, and the resulting statistics will not reveal the original quantum entanglement through violations of Bell's inequalities.
In the inflationary case, it may be that the time and rate of decoherence for inflationary fluctuations puts theoretical constraints on the parameter space of models that may allow for Bell inequality violations.
It would be interesting to pursue this using by perturbatively evolving the wave functional as we have done here.

There is also the question of whether any observable effect could arise from modes that do not decohere during inflation.
Because of the delay in decoherence, some short-wavelength modes may not fully decohere by the end of inflation.
However, 
it is likely that the presence of a bath of hot particles during or after reheating would decohere these modes as they re-entered the horizon. 
It would be valuable to better understand quantum-to-classical behavior in this context to see if any observable remnants could exist.

Lastly, we emphasize that while the decay of $\dot{\zeta}$ and the squeezing of the quantum state ensure that observations will appear classical, they does not explain how a single configuration of $\zeta(\x)$ is drawn stochastically from the distribution $|\Psi|^2$, that is how a measurement is made. This is described by the dynamical process of quantum decoherence, or equivalently the suppression of off-diagonal components of the reduced density matrix.
(Of course, we are not really addressing the measurement problem itself -- the question of why the world we observe is one particular outcome from the distribution $|\Psi|^2$ -- but the vanishing of off-diagonal components of the reduced density matrix for $\zeta$ does put this ``measurement'' on the same footing as more familiar quantum measurements.)

In particular, it is necessary to understand when decoherence occurs in order to determine whether super-Hubble modes can be viewed as giving stochastic kicks to the local value of the inflaton or scale factor \cite{Boddy:2014eba}, 
potentially allowing for a regime of slow-roll eternal inflation (which assumes a stochastic treatment of long-wavelength modes).
Assuming $\zeta$ can be treated stochastically, eternal inflation may occur when $H/\sqrt{\eps}\Mp\sim1$, that is when $\<\zeta^2\>$ becomes $\O(1)$ \cite{Creminelli:2008es}.
While there may be a longer delay in $e$-folds before modes behave stochastically if $\eps$ and $\eta$ are extremely small, decoherence will still occur after a finite amount of inflation. Thus, the picture of slow roll eternal inflation is unaltered. (Note also that when $\eps=0$ we are in exact de Sitter space, and our calculation does not apply.)
In this regime, the decohering effect of extremely IR modes becomes significant (\S \ref{sec:IR}).

\subsection{Future Directions}
\label{sec:future}

Several directions for future work are possible:
\begin{itemize}
\vspace{-0.15cm}
\item \textit{Quantum Origin of Perturbations.} As noted in \S \ref{sec:obs_q_grav}, it is interesting to ask how inflationary decoherence affects or constrains the observability of signatures of the quantum origin of cosmological perturbations in scenarios beyond single-field that may produce Bell inequality violating statistics.
\vspace{-0.2cm}
\item \textit{Squeezed-Limit Behavior.} We made some comments on the relation to soft-limit consistency relations in cosmology in \S \ref{sec:cons_rel}; it will be important to better understand the relevance of these soft-limit theorems or Ward identities for decoherence, which are a consequence of diffeomorphism invariance. Perhaps the quadratic and cubic parts of the phase of the wave functional, which lead to squeezing and decoherence respectively, are fixed in relation to one another in the same way. 
\vspace{-0.2cm}
\item \textit{Position Space vs. Fourier Space.} Our computation is in Fourier space, and thus obscures the spatially local process of decoherence in real-space. It would be interesting to carry out a similar computation, with a decomposition of the degrees of freedom of $\zeta$ into basis functions that are localized in both real space and Fourier space, capturing both the spatially local dynamics and the coupling of scales. The natural choice may be Gaussian wavepackets.
As a consequence of spatial locality, only overlapping wavepackets would be significantly coupled.
The role of short scale fluctuations in decohering large scales would perhaps be seen in the contribution to decoherence
of large wave packets due to narrower (overlapping) wave packets.  Such an approach might shed light on the relevance of soft-limit consistency relations for decoherence of a long-wavelength background.
It would also be interesting to understand the relation of decoherence from interactions to the decoherence (in the free theory) due to spatial gradient coupling, which was studied in \cite{Sharman:2007gi}.
\vspace{-0.2cm}
\item \textit{Dynamics in Phase Space.} It would be interesting to better understand the behavior of the quantum state for a given mode after decoherence. After the ``measurement'' of a classical value, the quantum state for a long-wavelength mode will be sharply peaked in both the configuration $(\zeta)$ and conjugate momentum $(\pi^{(\zeta)})$ directions in phase space.\footnote{As a result of the squeezing of the state, each classical value $\zeta_\k$ corresponds to a classical value for the conjugate momentum \cite{Kiefer:2008ku}.} It would be interesting to understand the dynamics of this state, including coherence lengths and squeezing at late times, and stochastic effects from the continuous monitoring of the short-scale environment, by studying the evolution of the associated Wigner function.\footnote{I thank Jess Riedel for discussions on this point.} It would be interesting to better understand the interplay between quantum decoherence and squeezing in determining the quantum state, as both are artifacts of the exponential expansion.
\vspace{-0.2cm}
\item \textit{Post-Inflation Decoherence.} It may be possible to follow the evolution of the quantum state into a (p)reheating era, for example with a simple $m^2\phi^2$ potential, by perturbatively evolving the wave functional under the influence of weak couplings. Optimistically, the degree of coherence or squeezing of short modes that do not decohere by the end of inflation might affect observable quantities.
\vspace{-0.2cm}
\item \textit{Decoherent Histories and Classicality.} Here we have approached decoherence by studying the dynamics of the reduced density matrix. The decoherent or consistent histories formalism (see \cite{Halliwell:1994uz,GellMann:1992kh,Hartle:1992as}, for example) could provide a complementary view of decoherence during inflation due to gravitational couplings, in the language of the decoherence functional.
Our study suggests that the appropriate prescription for coarse graining of decoherent histories (paths in field space) would be to smooth on a physical scale $\sim1/[H\cdot(\rm coupling \ \rm strength)]$.
\vspace{-0.2cm}
\item \textit{Quantum Darwinism.} Our computation also shows the proliferation of records from a long mode carried in the phases of short modes.
It would be interesting to study the application of Quantum Darwinism \cite{qdarwin,qdarwin2}, in which quantum states (in this case classical configurations of $\zeta$) survive by propagating records into their environment, to inflationary fluctuations. We note that in the present case, a long mode decoheres before individual short modes make records of it. The decoherence rate becomes large due to the sum over many environment modes, so it is only in the combined state of all these modes that a record is made.
\vspace{-0.2cm}
\item \textit{Entropy and Entanglement.} Lastly, it would be interesting to better understand the role of gravity in producing a classical spacetime with inhomogeneities in the context of quantum cosmology and quantum gravity. 
The pure-to-mixed transition from gravitational couplings in a quasi- de Sitter space, which we have studied here, will lead to the production of entropy, a much discussed aspect of de Sitter space. It would be interesting to study the entanglement entropy from mode coupling, $S=-{\rm Tr}(\hat{\rho}_{\rm R}\ln\hat{\rho}_{\rm R})$, using the reduced density matrix which we have computed.
\end{itemize}

\section*{Acknowledgments}

I am grateful for illuminating conversations with Niayesh Afshordi, Nishant Agarwal, Suddhasattwa Brahma, Cliff Burgess, Rich Holman, Matt Johnson, Jess Riedel, Sarah Shandera, and Kendrick Smith, and thank Jess Riedel and Suddhasattwa Brahma for comments on an earlier draft.
Research at Perimeter Institute is supported by the Government of Canada through Industry Canada and by the Province of Ontario through the Ministry of Research \& Innovation.

\appendix

\section{Computation of Wave Functional Evolution}
\label{app:WF}
In this Appendix we compute the evolution of the nonlinear part -- Eq. \eqref{F_solution} -- of the wave functional $\Psi[\zeta]$. We reproduce the $\zeta(\del\zeta)^2$ three-point function in the late-time limit, and show that the Hubble scale acts as a cutoff for the imaginary part ${\rm Im}\F$ which generates decoherence, with sub-horizon contributions dropping to zero quickly.

Because the environment is short-wavelength modes of $\zeta$ itself, we set $f_\E=f_\zeta$, which leads to Eq. \eqref{alpha_def} taking the form
\be\label{alpha_zzz}
\alpha_{k,k',q}(\tau) = - \tau^2\(\frac{1-\frac{i}{k\tau}}{1+k^2\tau^2}k^3 + \text{2 perms.}\),
\ee
which leads to
\be\label{alpha_int}
i \int^\tau d\tau'\alpha_{k,k',q}(\tau') = \[ -i k\tau - \ln(1 - ik\tau) \] + \text{2 perms.} \\
\ee
Plugging this into Eq. \eqref{F_solution} with $\tau_0\rarr-\infty$, we have
\be\label{F_zzz_exact}
\F_{\k,\k',\q}(\tau) = i \int_{-\infty}^\tau\frac{d\tau'}{H\tau'} \tHint_{\k,\k',\q}(\tau') e^{ik_t(\tau'-\tau)}\frac{1-ik\tau'}{1-ik\tau}\frac{1-ik'\tau'}{1-ik'\tau}\frac{1-iq\tau'}{1-iq\tau},
\ee
where $k_t\equiv k+k'+q$.
From the source $\tHint$, given in Eq. \eqref{source_zzz}, we have a factor of $1/\tau'$, leading to the time integral
\be
i \int_{-\infty}^\tau \frac{d\tau'}{\tau'^2} e^{ik_t(\tau'-\tau)}(1-ik\tau')(1-ik'\tau')(1-iq\tau') = \frac{-i}{\tau} - \frac{kk'+kq+k'q}{k_t} - \frac{kk'q}{k_t^2}(1-ik_t\tau),
\ee
where we have discarded the oscillatory lower limit by deforming the contour into the imaginary plane. The late-time limit of the real part of $\F$ is then
\be
\left.{\rm Re}\F_{\k,\k',\q}\right|_{\tau\rarr0}= - \frac{\eps\Mp^2}{6H^2}(\eps+\eta)(k^2+k'^2+q^2)\(-k_t + \frac{kk'+kq+k'q}{k_t} + \frac{kk'q}{k_t^2} \).
\ee
Note that this reproduces the momentum dependence of the part of the late-time three-point function,
\be
\<\zeta_\k\zeta_{\k'}\zeta_\q\> = \int\D\zeta(\zeta_\k\zeta_{\k'}\zeta_\q)|\Psi[\zeta]|^2,
\ee
generated by the $\zeta(\del\zeta)^2$ interaction -- see Eq. $(4.35)$ of \cite{Chen:2006nt} -- as it should.\footnote{The same time integral arises in the $\zeta\gamma\gamma$ and $\zeta\zeta\gamma-$type couplings between scalar and tensor modes; see, e.g., Eq. $(4.11)$ of \cite{Maldacena:2002vr}. The imaginary part which we study here is the part typically discarded when computing correlation functions, which only depend on diagonal elements of the density matrix.}

For the imaginary part, we consider the limit where the system mode is superhorizon, $|q\tau|\ll1$, but the environment modes $k\approx k'\gg q$ may be in either the subhorizon or superhorizon regime. From Eq. \eqref{F_zzz_exact},
\be\label{ImF_zzz_qeta}
\left.{\rm Im}\F_{\k,\k',\q}(\tau)\right|_{|q\tau|\ll1} = - \frac{\eps\Mp^2}{6H^2}\frac{1}{\tau}(\eps+\eta)\frac{k^2+k'^2+q^2}{(1+k^2\tau^2)(1+k'^2\tau^2)} + \O(q\tau).
\ee
When the environment modes are in the subhorizon regime, $|k\tau|\gtrsim1$, ${\rm Im}\F_{\k,\k',\q}$ is suppressed as $(k\tau)^{-5}$.
This ensures that the integral over the environment in $\Gamma_{\rm deco}$ is convergent in the UV (as shown in the main text), with the Hubble scale $aH=-1/\tau$ effectively acting as a UV cutoff. (The real part ${\rm Re}\F$, on the other hand, leads to UV divergences in loop integrals for correlation functions, and needs to be renormalized.)
The dominant contribution in the integral over the environment comes from the squeezed limit $k,k'\gg q$, for which we recover Eq. \eqref{ImF_zzz}. The subdominant contributions from super-Hubble environment modes $k\lesssim q$ lead to the correction described in \S \ref{sec:IR}.

\section{Computation for Speed of Sound $c_s\leq1$}
\label{app:cs}

In this Appendix we generalize the computation of \S \ref{sec:zzz} and Appendix \ref{app:WF} to include a nontrivial speed of sound $c_s<1$. From the quadratic action, Eq. \eqref{S2zeta}, we obtain the conjugate momentum,
\be
\pi_\k^{(\zeta)} = \frac{2\eps\Mp^2}{c_s^2}a^3\dot{\zeta}^*_\k,
\ee
and the free Hamiltonian,
\be
H_\zeta = \frac{1}{2}\int\frac{d^3\k}{(2\pi)^3}
\(\frac{c_s^2}{2\eps\Mp^2}\frac{1}{a^3} \pi^{(\zeta)}_\k\pi^{(\zeta)}_{-\k}
+ c_s^2 k^2 \frac{2\eps\Mp^2}{c_s^2} a \zeta_\k\zeta_{-\k}\),
\ee
which gives us
\be\label{f_zeta_cs}
f_\zeta(\tau) = -\frac{c_s^2}{2\eps\Mp^2}H^3\tau^3,
\ee
as well as the Gaussian wave functional evolution,
\be\label{AzetaBD_cs}
A_\zeta(k,\tau) = 2k^3 \frac{\eps c_s \Mp^2}{H^2} \frac{1 - \frac{i}{c_s k\tau}}{1+c_s^2k^2\tau^2}.
\ee
This leads to the simple replacement $\alpha_{k,k',q}\rarr\alpha_{c_sk,c_sk',c_sq}$. As a consequence, the maximum wavenumber will instead by $\kmax(a)=aH/c_s$, so modes that have just crossed the sound horizon will be the leading source of decoherence.

In the case of more general single field models of inflation with a nontrivial speed of sound, new self-interactions of $\zeta$ can arise, in particular a large $\dot{\zeta}^3$ interaction. As we have noted, though, such terms will not induce decoherence. More importantly for our purposes, $c_s<1$ leads to an enhancement of the $\zeta(\del_i\zeta)^2$ term \cite{Chen:2006nt}:
\be
\L_{\zeta\zeta\zeta}\rarr \eps\frac{1-c_s^2}{c_s^2} a\zeta(\del_i\zeta)^2 + \O(\eps,\eta)^2,
\ee
which is now larger by a factor of $1/\eps c_s^2$.\footnote{The term going like $\eta$, on the other hand, is not enhanced by $1/(\rm slow \ \rm roll)$.} 
When calculating the decoherence rate, which scales as $({\rm Re}A_\zeta)^{-2}$, we gain an additional factor of $1/c_s^2$ from Eq. \eqref{AzetaBD_cs}, as well as a factor of $1/c_s$ from integrating over environment modes up to the sound horizon $\sim aH/c_s$. Combining these factors we make the replacement
\be
(\eps+\eta)^2 \rarr \frac{(1-c_s^2)^2}{c_s^7}
\ee
in Eq. \eqref{d_boxed_final}. Because the nontrivial speed of sound enhances the interaction, the power spectrum for the environment, and the range of environment modes which are sensitive to the system mode, the decoherence rate is very sensitive to it: $\Gamma_{\rm deco}\propto 1/c_s^7$.
Decoherence now happens
\be
N^{(c_s<1)}_{\rm deco}\simeq\frac{1}{3}\ln(c_s^7/\Delta_\zeta^2)
\ee
$e$-folds after horizon crossing, with the decohering environment peaking at $c_s k_{\rm env} \sim aH$.
For $c_s=1/2$, for example, decoherence occurs after roughly $5$ $e$-folds.
For $c_s=1/20$, close to the \textit{Planck} bound from constraints on non-Gaussianity \cite{Ade:2015ava}, it could occur around the time of horizon crossing.

\bibliographystyle{JHEP}
\bibliography{deco_bib}

\end{document}